\documentclass[twocolumn]{aastex631}
\usepackage{silence}
\usepackage{graphicx}
\usepackage[caption=false]{subfig}
\usepackage{booktabs}
\usepackage{stmaryrd}
\usepackage{xcolor}
\usepackage{natbib}
\setcitestyle{authoryear,open={(},close={)}}

\usepackage{float}
\usepackage{color,soul}
\usepackage{wrapfig}
\usepackage{bm}
\usepackage{amsmath}

\usepackage{fancyhdr}
\usepackage{wasysym}
\usepackage{CJKutf8}

\newcommand{\md}{$M_{\rm d}$}
\newcommand{\mdeb}{$M_{\rm d,EB}$}

\begin{document}

\title{Simulation-Informed Dust Masses in Typical $6<z<14$ JWST Galaxies}
\author[0000-0002-6149-8178]{Jed McKinney}
\altaffiliation{NASA Hubble Fellow}
\affiliation{Department of Astronomy, The University of Texas at Austin, Austin, TX 78712, USA}
\affiliation{Cosmic Frontier Center, The University of Texas at Austin, Austin, TX 78712, USA}

\author[0009-0008-7017-5742]{Dhruv T. Zimmerman}
\affiliation{Department of Astronomy, University of Florida, 211 Bryant Space Sciences Center, Gainesville, FL 32611 USA}
\affiliation{Department of Astronomy, The University of Texas at Austin, Austin, TX 78712, USA}
\affiliation{Cosmic Frontier Center, The University of Texas at Austin, Austin, TX 78712, USA}

\author[0000-0002-7064-4309]{Desika Narayanan}
\affiliation{Department of Astronomy, University of Florida, 211 Bryant Space Sciences Center, Gainesville, FL 32611 USA}
\affiliation{Cosmic Dawn Center (DAWN), Niels Bohr Institute, University of Copenhagen, Jagtvej 128, København N, DK-2200, Denmark}

\author[0000-0002-0930-6466]{Caitlin M. Casey}
\affiliation{Department of Physics, University of California Santa Barbara, Santa Barbara, CA, USA}

\author[0000-0002-8984-0465]{Julian B. Mu\~noz}
\affiliation{Department of Astronomy, The University of Texas at Austin, Austin, TX 78712, USA}
\affiliation{Cosmic Frontier Center, The University of Texas at Austin, Austin, TX 78712, USA}

\author[0000-0002-5653-0786]{Paul Torrey}
\affiliation{Department of Astronomy, University of Virginia, 530 McCormick Road, Charlottesville, VA 22904}
\affiliation{Virginia Institute for Theoretical Astronomy, University of Virginia, Charlottesville, VA 22904, USA}
\affiliation{The NSF-Simons AI Institute for Cosmic Origins, USA}

\author[0000-0002-1253-2763,gname=Hui,sname=Li]{\begin{CJK*}{UTF8}{gbsn}Hui Li (李辉)\end{CJK*}}
\affiliation{Department of Astronomy, Tsinghua University, Beijing, 100084, People's Republic of China}

\author[0000-0003-3816-7028]{Federico Marinacci}
\affiliation{Dipartimento di Fisica e Astronomia ``Augusto Righi'', Universit\`a di Bologna, via Piero Gobetti 93/2, I-40129 Bologna, Italy\\}
\affiliation{INAF, Osservatorio di Astrofisica e Scienza dello Spazio di Bologna, via Piero Gobetti 93/3, I-40129 Bologna, Italy\\}

\begin{abstract}
Direct detections of dust in typical $z>8$ galaxies remain untenable, despite the presence of dust suggested by rest-frame UV/optical spectro-photometry. Informed by cosmological zoom-in simulations where dust particles evolve on-the-fly, we present a novel method for inferring dust masses in high-redshift galaxies using stellar half-light radii and UV/optical attenuation. Our methods reproduce observed trends at lower redshift, ultimately diverging from Milky Way-based calibration factors by up to $\sim1.5$ dex. We apply this method to observed JWST galaxies from $6<z<14$, and find dust masses in the range $M_{\rm dust}\sim10^3-10^5\,M_\odot$ that are fully consistent with upper limits from ALMA, and suggest an early dust enrichment period by supernovae followed by efficient grain destruction in the hot gas. Future observations may use our tabulated calibration to infer dust masses over a wide range of stellar masses and redshifts. 
\end{abstract}

\section{Introduction\label{sec:intro}}
Dust in the interstellar medium of galaxies follows from star formation as heavy elements formed during stellar nucleosynthesis condense into molecules. Significant reservoirs of dust may therefore be coincident with the abundance of UV-bright galaxies at $z>10$ discovered by \textit{JWST} \citep[e.g.,][]{Cullen2024,Morales2024,Topping2022,Austin2024}, depending on how efficient dust production can be. On one hand, the Atacama Large Millimeter Array (ALMA) has shown that $\sim10^8\,M_\odot$ worth of dust can accumulate just $\sim700$ Myr into galaxy formation \citep[e.g.,][]{watson_dusty_2015,Hashimoto2018,Hashimoto2019,bakx_accurate_2021,Fudamoto2021,endsley_radio_2022,Ferrara2022,Bouwens2022,McKinney2023,Manning2025}; 
however, systematic and direct confirmation of dust at earlier times remains elusive despite deep radio/sub-mm efforts \citep[e.g.,][]{Bakx2026}, with only one detection of dust continuum beyond $z>8$ at $z=8.3$ \citep{Tamura2019}.

The most accurate dust masses in high-redshift galaxies will always come from direct observations of thermal dust continuum along the Rayleigh-Jeans tail where dust temperature bias is mitigated and the spectrum is optically thin \citep{Hildebrand1983,Hughes1998,Scoville2014,Santini2014}. The upcoming ALMA Wideband Sensitivity Upgrade may make this more accessible beyond $z>8$, but ALMA has a relatively small field of view and generally does not probe the Rayleigh-Jeans tail with $z>6$ detections \citep{Sommovigo2022}. Far-infrared space missions like PRIMA\footnote{PRIMA: The PRobe far-Infrared Mission for Astrophysics, \url{https://prima.ipac.caltech.edu/}} might improve high$-z$ dust masses by mitigating temperature degeneracies, depending on the sensitivities reached in its deepest surveys.
Compounded by the integration times needed to plausibly detect dust continuum at very high redshift, a statistical census of dust at $z>8$ is not presently achievable, certainly for typical galaxies fainter than the most extreme sources. 

Nevertheless, quantifying and characterizing dust is crucial for informing the corrections needed to accurately measure intrinsic luminosities \citep{mckinney_modeling_2025,Markov2024}. Moreover, dust sets radiative feedback mechanisms that regulate star-formation \citep{Tielens1985,Soliman2024}, which might play a role in facilitating the high star-formation efficiencies needed to reproduce observations at $z\gtrsim10$ \citep[e.g.,][]{Casey2024,Chworowsky2024}. If direct confirmation of dust in the sub-mm/radio is unachievable, rest-frame UV/optical inference on dust masses provides an alternative route. 

Dust absorbs UV/optical photons and attenuates light at these wavelengths (see \citealt{Salim2020} for a recent review). This provides a natural, albeit complicated, connection between deep JWST observations available at $z>8$ and the dust content in those galaxies. UV continuum slopes at $6<z<12$ hint at some degree of dust obscuration \citep{Cullen2024,Topping2024,Narayanan2025}, but quantifying that dust presently relies on assumptions calibrated in the Milky Way (e.g., \citealt{Casey2024lrds}, more on this later). New observations and simulations are showing this to be inaccurate as UV/optical attenuation law shapes inferred at $z>6$ look increasingly different from the local Universe \citep{Markov2024,mckinney_modeling_2025,Narayanan2025,shivaei_diversity_2025,casey_dust_2026}. 

In this paper we use simulations described in \cite{Zimmerman2026} and \cite{Narayanan2025}, that evolve dust on-the-fly, to calibrate a dust mass estimator at $z>8$. Section \ref{sec:sims} describes the simulations we use. Section \ref{sec:obs} presents observational data used to validate our method at lower redshifts, and samples that we apply our method to at $z>8$. We summarize the formalism behind calculating dust masses in Section \ref{sec:dustcalcs} and present our new calibration.  We present our results in Section \ref{sec:results}, discuss them in Section \ref{sec:discussion}, and present a summary of our conclusions in Section \ref{sec:conclusion}. 

\section{Simulation Data\label{sec:sims}}
We analyze the suite of cosmological zoom-in simulations described in \cite{Zimmerman2026}, which we refer to for a complete discussion of the simulations and the details of the dust model. In summary, the zoom-in simulations make use of the \textsc{arepo} \citep{Springel2010,weinberger_arepo_2020} hydrodynamic code and the \textsc{smuggle} physics model \citep{marinacci_simulating_2019} including dust as described in \cite{Narayanan2023} which builds on the \textsc{arepo} dust model of \cite{mckinnon_dust_2016,mckinnon_simulating_2017,mckinnon_simulating_2018} and the \textsc{SIMBA} dust model \cite{Li2019,Li2021}. The dust model includes 16 logarithmic bins in dust grain size that evolve on-the-fly in response to the gas and radiation conditions. This prescription gives rise to a flexibly evolving dust extinction law, which is post-processed with \textsc{powderday} \citep{narayanan_powderday_2021} to derive the emergent attenuation law. 

Our suite includes 107 halos with snapshots beginning at $z=15$, sampled every $\Delta z=0.25$ through the end of each run. For each snapshot we sample the galaxy over 8 viewing angles which we include in our analysis to marginalize over orientation. In addition to quantities like total stellar mass and total dust mass which are tracked in the simulation, the radiative transfer calculations in each snapshot/viewing angle provide forward-modeled attenuation laws as well as the total attenuation in the rest-frame optical ($A_V$) and UV ($A_{UV}$)  calculated at 5500\AA\ and 1500\AA\ respectively. We emphasize that these emerge from the dust model and do not represent fixed dust properties that otherwise significantly impact the attenuation law's normalization and shape. Between $z=5-15$, 50\% of the simulation snapshots have $A_V>0.05$, and 67\% have $A_V>0.01$ indicating some degree of dust attenuation. 

Throughout this work we average these simulations in bins of quantities like stellar mass, redshift, and $A_{\rm V}$. To account for the incomplete halo mass selection function implicit in cosmological zoom-in simulations, all of the averages reported in Tables and shown on Figures are calculated with weights on each snapshot based on their halo mass and the corresponding value of the halo mass function. We compute halo mass functions with \texttt{COLOSSUS} \citep{Diemer2018}, using the \cite{Watson2013} fitting function for virial halo masses. 

\subsection{Simulated galaxy sizes}

Defining a simulated galaxy's size will be important later on as $M_{\rm dust}\propto R^2$. 
In this paper, we choose to use the radius enclosing 50\% of the simulated stellar mass at a given redshift to best match an empirical half-light radius calculated in the rest-frame optical. Observed galaxy sizes from \textit{JWST}/NIRCam's long wavelength filters trace rest-frame optical light at $6<z<14$. Optical emission is dominated by lower mass stars that comprise the bulk of the total stellar mass \citep{conroy_modeling_2013}, and is therefore less sensitive to uncertain mass-to-light ratios in the UV. Adopting half-mass radii best reproduces the observed size-mass and size-redshift relation as discussed in Section \ref{sec:size_mass_consistency}.\footnote{Observed sizes depend on wavelength in ways that might be impacted by dust attenuation. Future work is needed to transform the radiative transfer runs into mock observations and pull out sizes as done in observations.} This choice also follows the formalism presented in \cite{casey_dust_2026} where stellar radii are used in favor of dust sizes as the former is practically measurable in large samples and out to high$-z$.

\section{Comparison Observations\label{sec:obs}}

\subsection{Star-forming galaxies at $z<7$}

\cite{casey_dust_2026} presents an analysis of the average UV/optical through far-IR/sub-mm properties of $\sim500,000$ galaxies in the COSMOS-Web field \citep{Casey2022,Shuntov2025,Franco2026}. Fitting multi-wavelength stacked photometry their sample includes, among other quantities, $M_{\rm dust}$, UV/optical attenuation, and galaxy sizes for $\log M_*/M_\odot>9.5$ out to $z=6$. We use these stacks to validate the behavior of our simulations at $z<7$ and connect the emergent picture of dust formation in the early Universe to the behavior of typical galaxies through to $z\sim2$. 

\subsection{Reference high-redshift catalogs\label{sec:obs:epochs}}
To further benchmark our simulations, and to test our novel dust mass calibration, we use the catalogs from  ``Observing the First Epochs of Galaxy Assembly and Reionization'' (EPOCHS), a meta-analysis of 11 deep JWST/NIRCam extragalactic fields homogeneously reduced to reach F444W $5\sigma$ depths of $28.7-30.2$ mag \citep{epochsI_Conselice2025_surveyoverview}. EPOCHS includes 1167 $6.5<z<18$ galaxies with physical properties from spectral energy distribution (SED) modeling as presented by \cite{epochs4seds}\footnote{The EPOCHS SED catalog was obtained on \url{https://github.com/tHarvey303/EpochsI}}. We use catalog parameters with the \texttt{zgauss} extension where the redshift was allowed to vary during the fits. In need of NIRCam/F444W half-light radii, we match EPOCHS to the morphological catalogs from the DAWN JWST Archive (DJA) as described in \cite{Genin2025}. Galaxy sizes obtained this way are fully consistent with the galaxy size-mass and size-redshift relations in EPOCHS presented by \cite{epochs6sizes}. 

\cite{epochs4seds} thoroughly explores the impact of various priors on galaxy properties inferred in EPOCHS. Most important to our work is the choice of dust attenuation law, and priors on its overall shape and normalization. As discussed in \cite{Zimmerman2026}, our simulations exhibit relatively flat UV/optical dust attenuation laws. The fiducial SED fits from \cite{epochs4seds} adopt a \cite{calzetti_dust_2000} prescription, which while not perfectly matched to our simulations is at least significantly less deviant than steeper models \citep{mckinney_modeling_2025}. The simulations on-average exhibit flatter attenuation laws in the UV than \cite{calzetti_dust_2000}, which could bias $A_V$ and stellar mass measurements low by as much as $0.5$ mag and $0.4$ dex respectively \citep{mckinney_modeling_2025}. It is beyond the scope of this work to re-model the EPOCHS catalog using our theoretically-derived dust attenuation models for a fully self-consistent comparison. 
For now we adopt the fiducial SED fits of \cite{epochs4seds}. 


\subsection{Size-Mass and Size-redshift scaling\label{sec:size_mass_consistency}}

\begin{figure}[t]
    \centering
    \includegraphics[width=\linewidth]{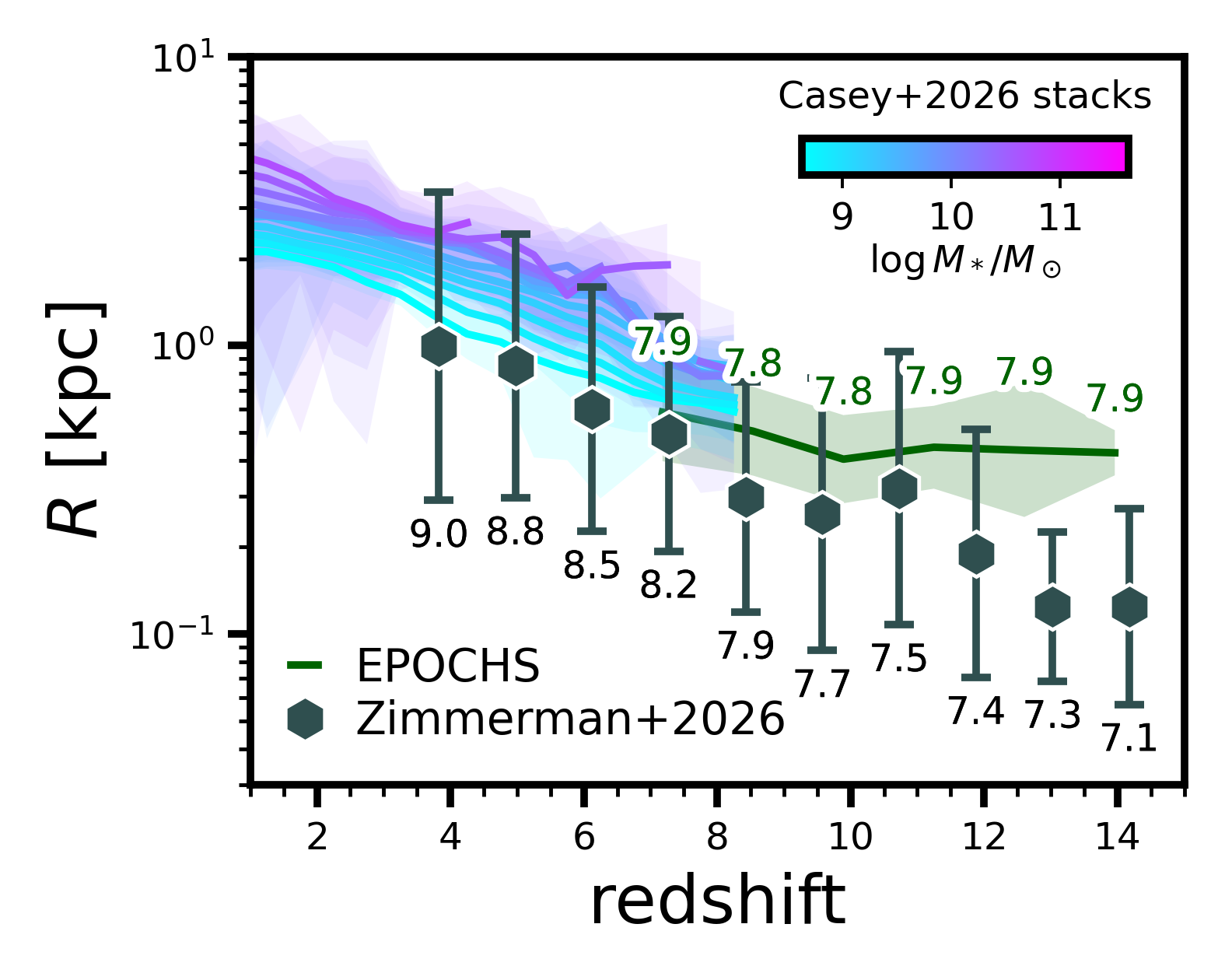}
    \caption{\textbf{Observed and simulated size-redshift relation.} The median stellar mass of our simulated galaxies is listed below each redshift bin. Galaxy stacks from \cite{casey_dust_2026} are shown for comparison at $z<8$, and from EPOCHS for $6.5<z<15$ \citep{epochs6sizes}. By $z\sim6$ our simulations overlap the stellar mass range of \cite{casey_dust_2026} where we reproduce the observed relation approaching $\log M_*/M_\odot\sim9$. Our simulations continue to track the sizes of observed galaxies of comparable stellar mass until $z\sim11$, after which they probe a systematically smaller mass regime than EPOCHS by $\sim0.4-0.7$ dex with accordingly more compact sizes. 
    }
    \label{fig:size_z}
\end{figure}

\begin{figure}[t]
    \centering
    \includegraphics[width=\linewidth]{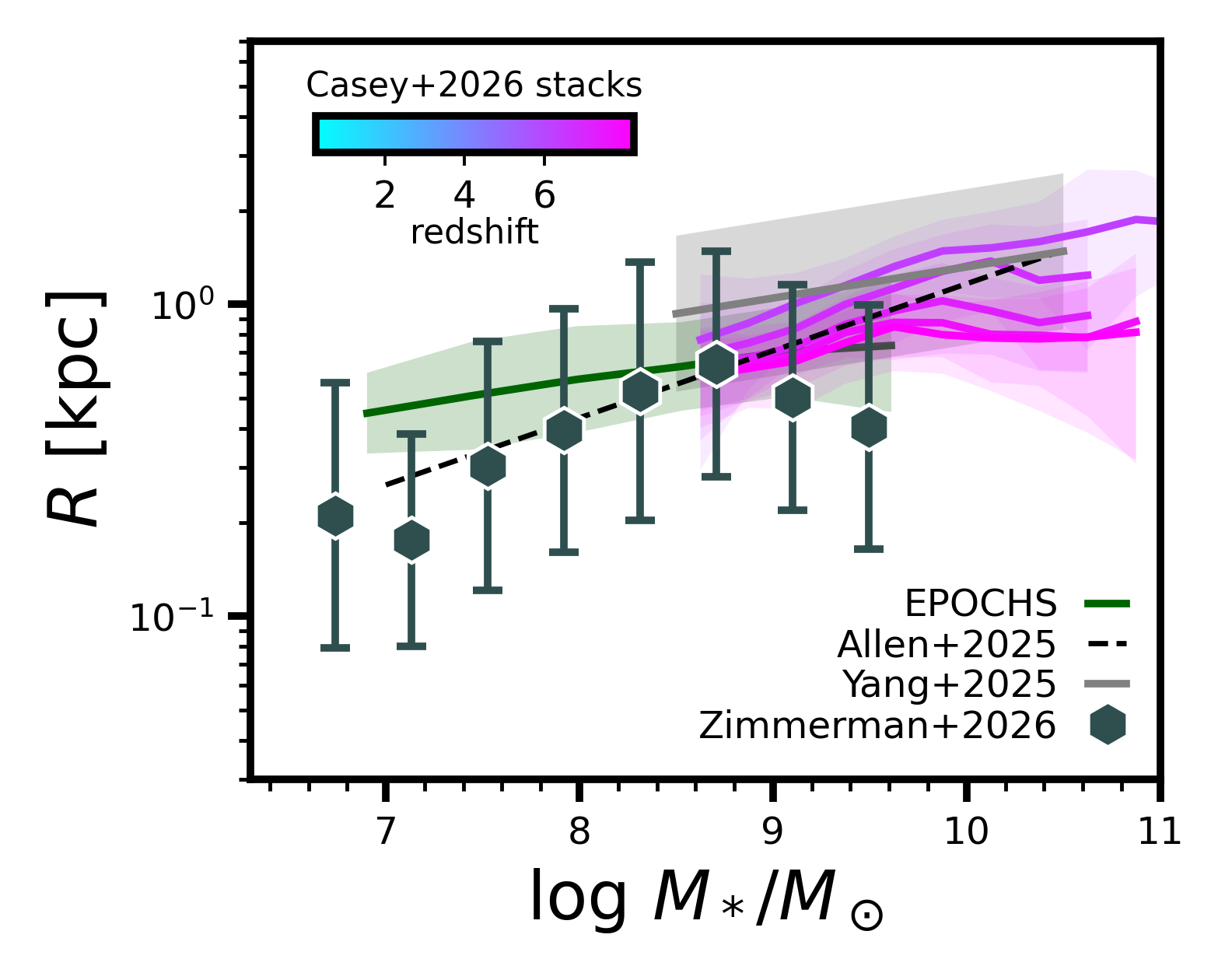}
    \caption{\textbf{Observed galaxy size-mass relation relative to our simulations at $6<z<9$}. We compare to the \cite{casey_dust_2026} stacks, EPOCHS \citep{epochsI_Conselice2025_surveyoverview,epochs6sizes}, and empirical trends from \cite{Allen2025} and \cite{Yang2025}. Our simulations overlap with observations of comparable stellar mass within the scatter, with some deviation at the lowest stellar masses relative to EPOCHS. We note that the \cite{Allen2025} relation is extrapolated below their mass limits.}
    \label{fig:size_mass}
\end{figure}

Figure \ref{fig:size_z} shows the evolution of half-mass sizes in our simulations as a function of redshift. Our simulated galaxies follow the increase in observed sizes from \cite{epochs6sizes} and \cite{casey_dust_2026} with redshift. Above $z\sim11$, our simulations approach stellar masses $\sim1$ dex below the mass limits of current observations, and as a result our median sizes evolve to be lower than redshift-matched observations. Reassuringly, our simulations reproduce the mass-size relation at $6<z<9$ as shown in Figure \ref{fig:size_mass}. Galaxy sizes are an important systematic when using optical attenuation to infer a dust mass \citep{Ferrara2024bm,casey_dust_2026}; however, our broad agreement with observed sizes as a function of both mass and redshift mitigates this concern. 


\section{Calculating Dust Masses\label{sec:dustcalcs}}
We refer to Section 2 of \cite{casey_dust_2026} for a thorough overview and methodological review behind the key concepts and assumptions connecting UV/optical attenuation with the mass and extent of dust. In this work we build off the formalism summarized therein, with the following key points worth reiterating:
\begin{enumerate}
    \item Dust sizes are assumed to evolve consistently with stellar radii, such that $R_{\rm dust}\approx R_{\rm stars}/\sqrt{2}$ due to centrally concentrated dust reservoirs \citep[e.g.,][]{Conselice2014,Mosenkov2019,casey_dust_2026}.
    \item The effects of star-dust geometry and dust grain properties through the dust mass absorption coefficient are marginalized over.
\end{enumerate}

For lack of direct dust mass measurements using far-infrared/sub-mm observations, we proceed to summarize two methods that are used to estimate $M_{\rm dust}$ using only UV/optical diagnostics. 
\newpage

\subsection{Dust Masses From Attenuation and Galaxy Sizes}

Assuming a uniform, foreground screen of dust, the monochromatic attenuation relates to the dust mass surface density ($\Sigma_{\rm dust}$) through:
\begin{equation}
    A_\lambda=C_\lambda\times\frac{\Sigma_{\rm dust}}{10^5\,M_\odot\,\rm kpc^{-2}}\,\,\,\rm mag
    \label{eq:av}
\end{equation}
where the pre-factor $C_\lambda$ captures the effects of dust properties and star-gas geometry \citep[e.g.,][]{casey_dust_2026}. It is useful to discuss the behavior of $C_\lambda$ in the $V$ band for historical and conventional purposes. Typically $C_V$ is fixed by scaling arguments anchored to the observed extinction-to-Hydrogen column density ratio $A_V/N_H=5.34\times10^{-22}\,{\rm mag\,cm^2\,H^{-1}}$ in the Milky Way \citep{Bohlin1978}, and canonical dust-to-Hydrogen and dust-to-gas mass ratios \citep{Draine2007,RemyRuyer2014}. Under these circumstances $C_V=0.74$ \citep{Aniano2012,Draine2014}, with some studies advocating for $C_V=0.18$ based on combined optical and far-infrared observations of nearby and distant galaxies \citep{Kreckel2013,Hodge2024}. \cite{casey_dust_2026} recently measured the redshift and stellar mass dependence of $C_V$ using multi-wavelength galaxy stacks out to $z\sim7$, finding that $C_V$ approaches $\sim0.1$ towards higher stellar mass and redshift. 

Equation \ref{eq:av} can be re-written using $\Sigma_{\rm dust}=M_{\rm dust}/2\pi R_{\rm dust}^2 $ such that
\begin{equation}
    M_{\rm dust}=\frac{2\pi\times10^5}{C_V}\times\frac{A_V }{{\rm mag}}\times\frac{R_{\rm dust}^2}{{\rm kpc^{2}}}\,\,\,M_\odot
    \label{eq:mdust}
\end{equation}
which provides an avenue for estimating dust masses from rest-frame optical measurements, namely the total attenuation and assuming that the extent of optical light and dust are comparable. Note that the factor of $2$ in the denominator of $\Sigma_{\rm dust}$ accounts for dust behind the illuminating source in addition to the foreground screen that gives rise to any observed dust extinction. Equations \ref{eq:av} and \ref{eq:mdust} can be expressed as a function of wavelength, and re-written in terms of $C_{\rm UV}$ and the UV attenuation $A_{\rm UV}$, which will prove useful for the highest redshift galaxies lacking rest-frame optical measurements, with the caveat that UV-only inference on the total attenuation is significantly more uncertain. While the UV can be the most impacted by attenuation, observed UV continuum slopes are degenerate between the shape of the dust attenuation law, the unattenuated stellar spectrum, and nebular continuum \citep{Narayanan2025,Katz2025}. If UV attenuation is gray as suggested by our models and other observations \citep{Markov2024,shivaei_diversity_2025,mckinney_modeling_2025,Zimmerman2026}, then the shape of the UV spectrum alone could be insensitive to dust along the line of sight.

\begin{figure}
    \centering
    \includegraphics[width=\linewidth]{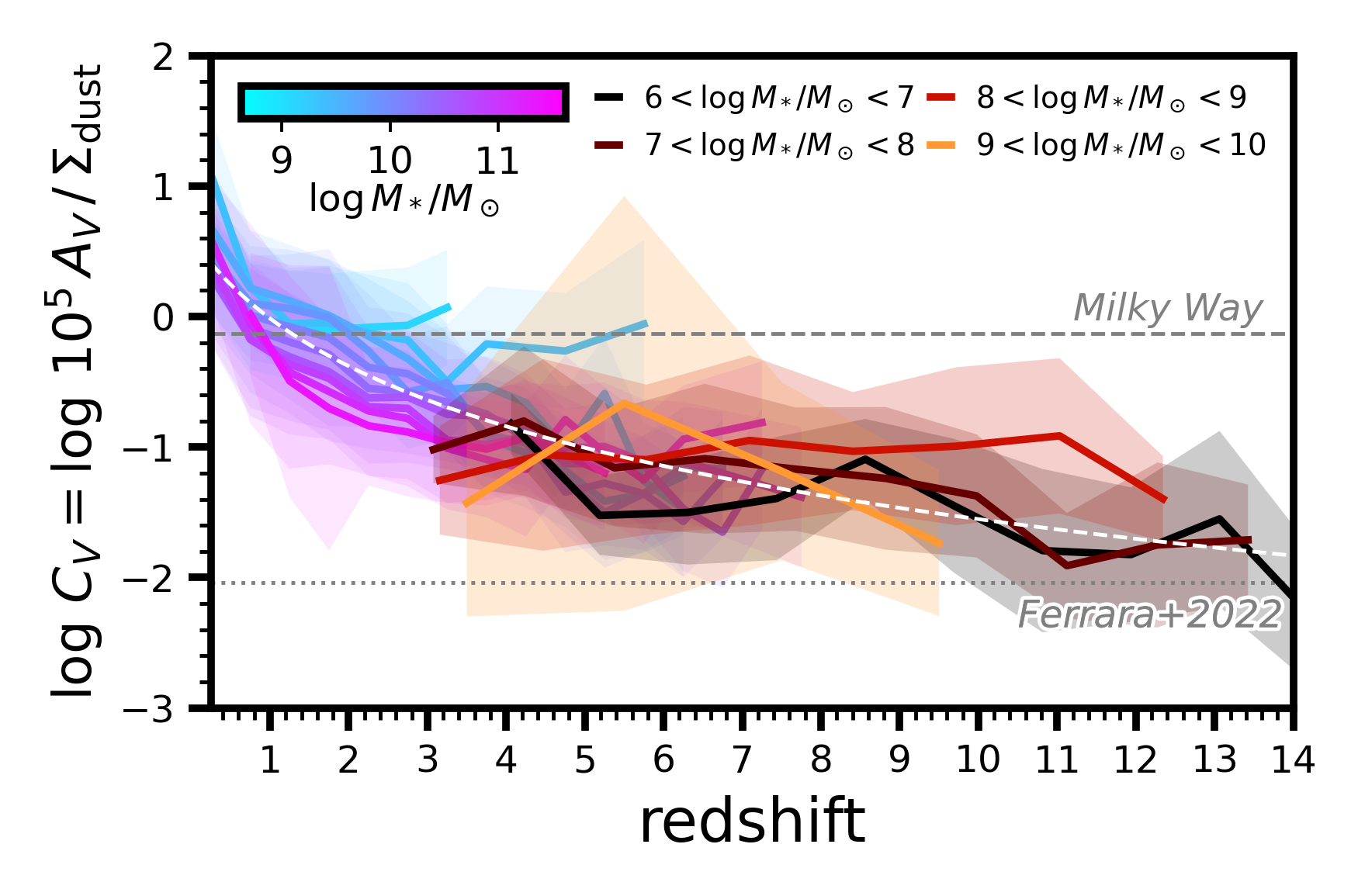}
    \caption{
    \textbf{Conversion factors between dust mass surface density and observed optical attenuation.} 
    Our simulations are shown as shaded regions in bins of stellar mass for snapshots with $A_V>0.01$. 
    We compare against \cite{casey_dust_2026} who measured $C_{\rm V}$ in galaxy stacks. The horizontal grey dashed line indicates the Milky Way value from \cite{Draine2014}, and the dotted grey line shows the theoretical value from \cite{Ferrara2022} calibrated for UV-bright $z\sim7$ galaxies. Our simulations broadly reproduce the observations where they overlap in redshift and stellar mass. Observations can not constrain $C_V$ statistically beyond $z\sim6$ below $\log M_*/M_\odot\sim11$, where our simulations evolve towards values lower than the Milky Way by $\sim1-1.5$ dex, approaching the estimate from \cite{Ferrara2022}. The white dashed line shows a joint fit to the observations and simulations from $z=0-14$, given in Equation \ref{eq:cvz}. 
    }
    \label{fig:logcv}
\end{figure}

\begin{deluxetable}{lccc}
\tablecaption{Halo-mass-function-weighted average $\log C_{\rm V}$ and $\log C_{\rm UV}$ by Stellar Mass and Redshift\label{tab:CV_CUV}}
\tablehead{
\colhead{Stellar Mass} & \colhead{Redshift} & \colhead{$\langle\log C_{\rm V}\rangle$} & \colhead{$\langle\log C_{\rm UV}\rangle$}
}
\startdata
$\log M_*/M_\odot<7$ & $2-4$ & $-0.52\pm0.23$ & $-0.24\pm0.22$ \\
 & $4-6$ & $-1.33\pm0.26$ & $-0.90\pm0.19$ \\
 & $6-8$ & $-1.04\pm0.48$ & $-0.72\pm0.57$ \\
 & $8-10$ & $-0.89\pm0.36$ & $-0.60\pm0.35$ \\
 & $10-12$ & $-1.45\pm0.63$ & $-1.11\pm0.69$ \\
 & $12-14$ & $-1.44\pm0.71$ & $-1.13\pm0.63$ \\
$7<\log M_*/M_\odot<8$ & $2-4$ & $-0.66\pm0.29$ & $-0.11\pm0.36$ \\
 & $4-6$ & $-0.77\pm0.54$ & $-0.40\pm0.55$ \\
 & $6-8$ & $-0.83\pm0.53$ & $-0.56\pm0.56$ \\
 & $8-10$ & $-0.95\pm0.51$ & $-0.67\pm0.53$ \\
 & $10-12$ & $-1.36\pm0.53$ & $-1.03\pm0.52$ \\
 & $12-14$ & $-1.34\pm0.47$ & $-0.97\pm0.46$ \\
$8<\log M_*/M_\odot<9$ & $2-4$ & $-0.98\pm0.42$ & $-0.77\pm0.51$ \\
 & $4-6$ & $-0.79\pm0.69$ & $-0.62\pm0.74$ \\
 & $6-8$ & $-0.63\pm0.61$ & $-0.37\pm0.65$ \\
 & $8-10$ & $-0.72\pm0.48$ & $-0.49\pm0.51$ \\
 & $10-12$ & $-0.75\pm0.51$ & $-0.47\pm0.43$ \\
 & $12-14$ & $-1.00\pm0.41$ & $-0.79\pm0.40$ \\
$\log M_*/M_\odot>9$ & $2-4$ & $-1.34\pm0.58$ & $-1.20\pm0.64$ \\
 & $4-6$ & $-0.43\pm1.47$ & $-0.25\pm1.51$ \\
 & $6-8$ & $-0.92\pm0.70$ & $-0.71\pm0.76$ \\
 & $8-10$ & $-1.61\pm0.63$ & $-1.43\pm0.58$ \\
 & $10-12$ & $-1.05\pm0.13$ & $-0.91\pm0.11$ \\
 & $12-14$ & \nodata & \nodata \\
\enddata
\tablecomments{Weighted averages ($\pm1\sigma$ weighted standard deviation) computed from our simulated galaxy sample restricted to $\tau_V>0.01$, with each galaxy weighted by the halo mass function $dn/d\ln M$ at its halo mass and redshift.}
\end{deluxetable}

\subsection{Dust Masses From Attenuation and UV Luminosity\label{sec:dustcalcs:eb}}
An alternative method to estimating dust masses without far-infrared data is to assume energy balance and map observed UV luminosities to a total IR luminosity: 
\begin{equation}
    L_{\rm IR}=L_{\rm UV,\,\rm obs}\times(10^{0.4A_{\rm UV}}-1)
\end{equation}
The IR luminosity can be converted to a dust mass with a mass-to-light ratio for a given dust SED template. \cite{Draine2014} models are commonly used \citep[e.g.,][]{Boquien2019,Burgarella2025}, and are parameterized as a function of the minimum and maximum radiation field intensity that the dust grains are exposed to ($U_{\rm min}$, $U_{\rm max}$), as well as the mass fraction of Polycyclic Aromatic Hydrocarbons ($q_{\rm PAH}$), the smallest grains. Over the range commonly adopted when fitting the SEDs of $z>6$ \textit{JWST} targets ($U_{\rm max}=10^6$, $U_{\rm min}\in[0.1,25],$ \citealt{Berta2025}), capturing the extreme radiation conditions of early galaxies, the \cite{Draine2014} models have $L_{\rm IR}/M_{\rm dust}\approx3920\,L_\odot/M_\odot$ on-average. Thus, without any size information and assuming energy balance the dust mass can be estimated using only $L_{UV}$ and $A_{UV}$. This technique is known to over-estimate the true dust mass by factors of $\sim3$ with significant biases against dust temperature \citep{Santini2014,casey_dust_2026}, and should generally be avoided except to set a generous upper limit. Nevertheless, we introduce this method as a baseline comparison against our calibrations using Equations \ref{eq:av} and \ref{eq:mdust}.


\begin{figure*}
    \centering
    \includegraphics[width=\linewidth]{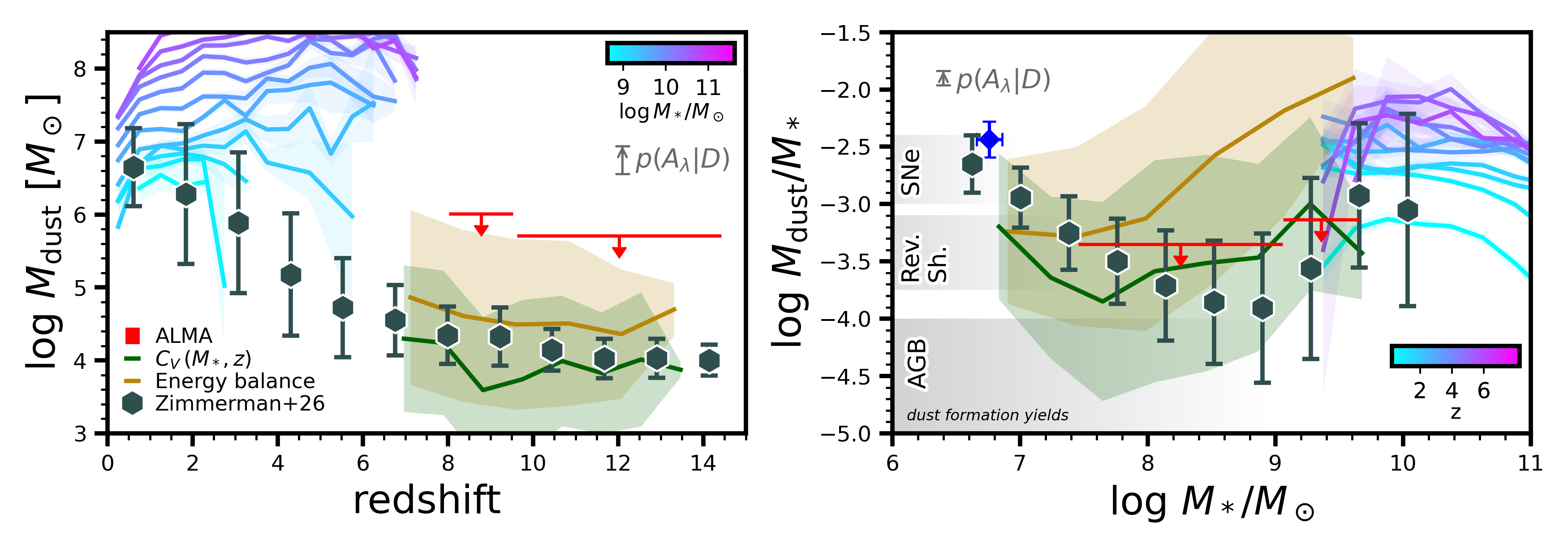}
    \caption{ \textbf{The evolution of total dust mass and dust-to-stellar mass ratios in simulations and observations from $\bf z\sim0-14$.}
    (\textit{Left})
    Average total dust mass in our simulations as a function of redshift, compared to the \cite{casey_dust_2026} stacks as well as observationally-derived trends described in Section \ref{sec:results}.  The median stellar mass is listed over each corresponding redshift bin. Our simulations (hexagons) and our fiducial empirical trend derived from our $C_V(M_*,z)$ method (green) agree within the respective scatter, indicating a typical dust mass of $\sim10^4\,M_\odot$ from $z=14$ to $z\sim6$. Relatively, energy balance arguments (gold) over-predict dust masses in this redshift range by $\sim1$ dex, but still falls below the deep ALMA upper limits at $z>8$ (red, \citealt{Bakx2026}).
    (\textit{Right}).
    Same samples as in the \textit{Left} panel, now showing their dust-to-stellar mass ratios vs. stellar mass for everything with $z>6$. We also show theoretical dust yield ranges as tabulated by \cite{Schneider2024}, associated with SNe production with and without reverse shock processing, as well as AGB stars. 
    Grey arrows and limits show the direction and 84th percentile range
    that the choice of attenuation law can induce: $p(A_\lambda|D)$. 
    Dust-to-stellar mass ratios in observed $z>6$ galaxies estimated with our $C_V$ methodology track the behavior of the simulations within the scatter. These remain consistent with the direct ALMA upper limits \citep{Bakx2026}, and appear to converge to high $M_{\rm dust}/M_*$ value inferred using our $C_V$ calibration with the lensed low-mass galaxy reported by \cite{Asada2026} at $z=5.5$ (blue diamond). 
    As in the \textit{Left} panel, energy balance arguments over-predict $M_{\rm dust}$ and would require sustained SNe formation yields with minimal destruction over the full mass range at $z>6$.
    }
    \label{fig:md}
\end{figure*}

\newpage
\section{Results\label{sec:results}}

\subsection{Calibrating Dust Masses as a Function of Size and Total Attenuation}

Given that our simulations track total dust masses, and that our radiative transfer calculations provide $A_V$ and $A_{\rm UV}$, we can read off $C_V$ and $C_{UV}$ for each snapshot and viewing angle following Equation \ref{eq:av}. We tabulate these quantities in Table \ref{tab:CV_CUV} in bins of stellar mass and redshift. In this calculation, we only consider snapshots with non-negligible dust attenuation ($A_V>0.01$), which accounts for $\sim65\%$ of all snapshots at $z>4$ and $\sim75\%$ at $z<4$. Figure \ref{fig:logcv} shows the evolution of $C_V$ in our simulations as a function of redshift and stellar mass. We recover the \cite{casey_dust_2026} trends by $z\sim7$ and track them down to $z\sim3$. At higher redshifts we see gradual evolution towards lower $C_V$, reaching values as low as $C_V\sim0.04$ at $z>10$ in the lowest mass galaxies with $M_*/M_\odot<10^7$. These approach the theoretical value from \cite{Ferrara2022} for Milky Way dust attenuation and a homogeneous sphere. The behavior of $C_{\rm UV}$ over mass and redshift is qualitatively the same. Fitting the \cite{casey_dust_2026} stacks and our simulations, with equal weights in each redshift bin, we find:
\begin{equation}
    \label{eq:cvz}
    \log C_{\rm V} = (0.61\pm0.26) -(2.08\pm0.34) \log(1+z)
\end{equation}

This redshift evolution in $C_{\rm V}$ can largely be explained by the fact that galaxy sizes are decreasing with redshift such that less dust is needed to produce the same $A_V$. This is counteracted by changes in the simulated dust attenuation laws as discussed in \cite{Zimmerman2026}. Preferential shattering of the smallest dust grains produces shallower attenuation laws, allowing for higher dust masses for fixed $A_V$. These relatively flat UV attenuation laws are in place in our simulations down to $z\sim5$, and could explain the $C_V$ floor of $C_V\gtrsim0.01$. 

\subsection{Applications to Observations}
We next use the $C_V$ grid across mass and redshift output by our simulations to estimate the dust masses of observed galaxies between $z=6-14$. As outlined in Section \ref{sec:obs:epochs} we use the EPOCHS catalog for this analysis, which for each galaxy yields a photometric redshift, stellar mass, and $A_V$ \citep{epochs4seds}, as well as a galaxy size \citep{epochs6sizes}. We calculate fiducial dust masses (\md) in EPOCHS using Equation \ref{eq:mdust} and the mass- and redshift-matched $C_V$ from Table \ref{tab:CV_CUV} for each source. Figure \ref{fig:md} compares these dust masses as a function of redshift and dust-to-stellar mass ratios against the simulations, the \cite{casey_dust_2026} stacks from $z\sim0-7$, and constraints from ALMA. As a check against this methodology, we also estimate dust masses as described in Section \ref{sec:dustcalcs:eb} from UV luminosities which instead relies on energy balance arguments and fixed dust templates. These are denoted \mdeb.

As shown in Figure \ref{fig:md}, the median fiducial dust masses estimated for $z\sim6-14$ galaxies agree with those in our simulations. \mdeb\ instead fall above the simulation median, with an upper envelope reaching $\sim1$ order of magnitude greater than simulated dust masses despite the comparable mass and size ranges. Both \md\ and \mdeb\ are below deep ALMA-derived upper limits at $z>8$ \citep{Bakx2026}. 

Notably, the observed dust masses estimated using our simulation-based priors on $C_V$ track the dust-to-stellar mass ratio (DTM) remarkably well for $\log\,M_*/M_\odot=7-9.5$. This captures the high $\log\rm DTM\approx-2.5$ at $\log\,M_*/M_\odot\sim7$ \citep{Zimmerman2026} and is bounded below by the dust yield of AGB stars \citep{Schneider2024}, suggesting that AGB dust production places a lower limit on the dust content of high-redshift galaxies. 
Figure \ref{fig:md} \textit{Right} demonstrates the failure of estimating dust from Energy Balance arguments alone; EPOCHS DTM ratios using \mdeb\ are $\sim0.5-1$ dex greater at the lowest stellar mass range, and up to $\sim1.5$ dex above ALMA limits and DTM(\md) by $\log M_*/M_\odot\sim9$. EPOCHS DTM ratios using our simulation-based priors are consistent with ALMA limits, on-average, over the full mass range. 

Modeling observations with a gray attenuation law that more closely matches those in our simulations can change the inferred $A_V$ by $+0.2$ mag on-average as well as stellar mass by $0.1$ dex on-average \citep{mckinney_modeling_2025}. This confers a $\sim0.5$ dex increase in $M_{\rm dust}$ inferred from Eq.\,\ref{eq:mdust} at most (median $+0.2$ dex), and $0.2$ dex maximal increase in the DTM. As illustrated on Figure \ref{fig:md} this is not enough to bring our simulation-informed dust masses into disagreement with the ALMA upper limits, and imposes a negligible systematic on the DTM. Re-measuring $M_{\rm dust}$ in EPOCHS from SED fits using attenuation laws more consistent with our simulations is likely to reduce any residual offset from our simulated dust masses, but is unlikely to change trends with DTM ratios.  

\section{Discussion\label{sec:discussion}}
As outlined in \cite{Zimmerman2026}, our simulations paint a picture in which high dust production rate from SNe at $z>8$ enrich the ISM efficiently with dust. This dust quickly mixes with gas of temperature and metallicity favorable to dust destruction through sputtering on time scales shorter than a depletion time, suppressing grain growth in the ISM and driving the DTM ratio down. By $z\sim5$, galaxies are sufficiently metal-enriched for gas cooling rates and the subsequent ISM temperatures to favor dust grain growth through coagulation, enabling the DTM to climb back up to canonical values observed in massive, dust-obscured galaxies at $z\sim1-3$ \citep{santini2010}. Alternatively, this climb might arise from AGB dust production turning on at $z\lesssim4$ \citep[e.g.,][]{schneider_dust_2014,Schneider2024,Burgarella2026}. The DTM relation then turns over at $\log M_*/M_\odot\sim10$ as stellar mass production outpaces net dust formation, given that dust is also removed via efficient gas conversion \citep{casey_dust_2026}.

\begin{figure}
    \centering
    \includegraphics[width=\linewidth]{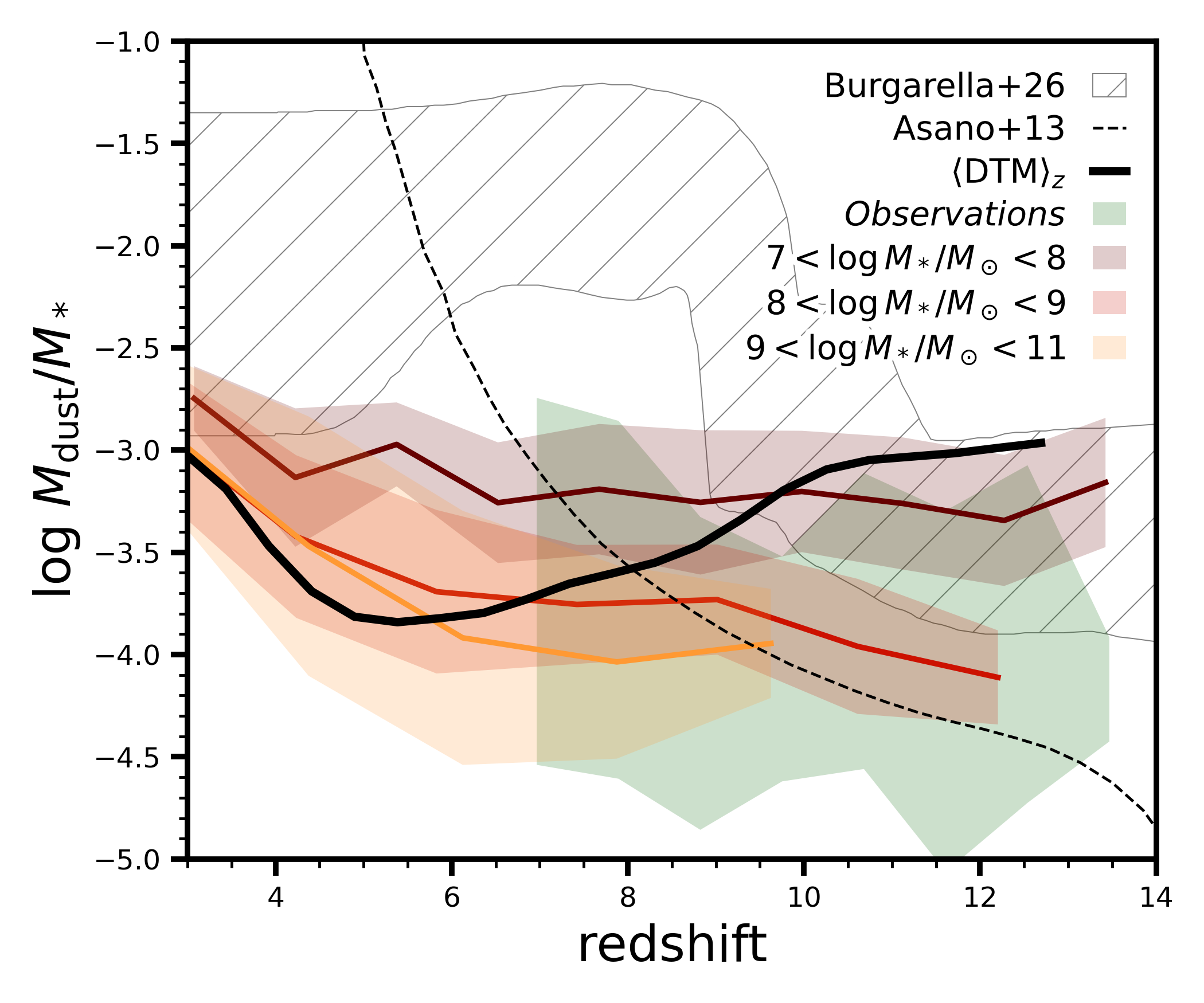}
    \caption{
    \textbf{Dust-to-stellar mass ratio evolution across models.} We compare our results to the fiducial model set of \cite{Burgarella2026}, and their adaptation of the \cite{Asano2013} model for $z_{\rm form}=15$. 
    Our simulations are shown in bins of stellar mass, and the solid black line indicates the mean evolutionary track for galaxies with $\log M_*/M_\odot\sim9.5$ at $z=2$.
    The simulation-informed observations (green) are consistent with a gradual increase in DTM seen in our more massive simulated galaxies, rather than the step function behavior predicted by \cite{Burgarella2026} arising from a sharp transition between SNe dust production and ISM grain growth.
    }
    \label{fig:dtmz}
\end{figure}

The general agreement between our simulations and observations on Figure \ref{fig:md} lends some support to this narrative. 
Remarkably, the DTM ratios inferred from observations using our methods agree with the simulations across $\sim3$ dex in stellar mass, including the rise in DTM towards the lowest masses. There is no evidence for significant stellar mass dependence in $C_V$ (Figure \ref{fig:logcv}); therefore, while \md\ is informed using the simulations, the observed stellar masses are fully independent. That being said, the population scatter on both observations and within the simulations themselves for fixed stellar mass is large. This could be reduced in future work that controls for orientation angle, and makes use of observations with improved SED constraints from JWST MIRI to anchor stellar masses, and/or NIRSpec for well-characterized UV/optical continuum and attenuation. 

Reproducing the observed dust content in massive $z\sim6$ galaxies has been a challenge for simulations without pushing dust production rates to their theoretical limits \citep[e.g.,][]{Choban2025}. Various models mix dust production and processing physics with a variety of assumed timescales and yields (see \citealt{Schneider2024} for a thorough review). Figure \ref{fig:dtmz} compares the DTM ratios in our simulations and observations (using our simulation-based priors at $z>7$) with semi-analytic models. \cite{Asano2013} couples the evolution of gas, stars, metals, and dust through supernova and AGB-star production, supernova destruction, and metallicity-dependent grain growth. \cite{Burgarella2026} assume that core-collapse supernovae produce the first dust grains, which seed later growth through accretion of gas-phase metals. Grain growth accelerates with metallicity, and competes with grain destruction, outflows, and removal during star-formation. Most notably, \cite{Burgarella2026} predict efficient growth near $ Z/Z_\odot\sim0.1$ and a sharp transition between dust-poor and dust-rich galaxies around $z\sim9$. 

In contrast to some semi-analytic models, our observed and simulated DTM ratios predict roughly flat DTM evolution with redshift until $z<4$ where efficient grain growth in the ISM turns on \citep{Zimmerman2026}. This behavior is more shallow than the dust growth models of \cite{Asano2013} for a galaxy with $z_{\rm form}=15$. We do not find evidence in our simulations for a sharp turnover in DTM around $z\sim9$ as in \cite{Burgarella2026}. Besides methodological differences between semi-analytic models and cosmological hydrodynamical simulations, the dust prescriptions have notable differences. Where \cite{Burgarella2026} assumes that grain growth scales $\propto Z^{-1}$, our simulations have grain accretion as the dominant growth channel with $\tau_{acc}\propto\rho_{\rm gas}^{-1}Z^{-1}T_{\rm gas}^{-1/2}$ \citep{Narayanan2023,Narayanan2025,Zimmerman2026}. 
Therefore, more detailed surveys of $\log M_*/M_\odot>8$ galaxies should prove constraining not only on the total dust content of high-redshift galaxies, but also the underlying dust formation prescription as these can produce different DTM evolutionary tracks for fixed stellar mass. 

\section{Conclusion\label{sec:conclusion}}
Using simulation-based priors to estimate dust masses from stellar half-light radii and optical attenuation for the first time, we find that JWST catalogs are consistent with $M_{\rm dust}\sim10^3-10^5\,M_\odot$ between $6<z<14$. These dust masses are supported by dust formation through SNe channels \citep{Zimmerman2026}, and in agreement with ALMA upper limits at $z>8$. 
We present our simulation-informed conversions for estimating $M_{\rm dust}$ from optical attenuation and galaxy stellar radii, tabulated as a function of stellar mass and redshift. These yield dust masses $\sim1$ dex lower than energy conservation arguments, as they encode the different dust properties emergent in the early Universe. Future observations using ALMA following the wide-band sensitivity upgrade, or  future far-IR space telescopes like PRIMA, might be able to verify these dust models, and the dust masses estimated here, by measuring dust-to-stellar mass ratios in $M_*=10^7-10^9\,M_\odot$ galaxies at $z>8$.

\begin{acknowledgments}
{\small
JM acknowledges the invaluable labor of the maintenance and clerical staff at our institutions, whose contributions make our scientific discoveries a reality. JM thanks NASA and acknowledges support through the Hubble Fellowship Program, awarded by the Space Telescope Science Institute, which is operated by the Association of Universities for Research in Astronomy, Inc., for NASA, under contract NAS5-26555. JM and DTZ acknowledge support from JWST-GO-03224, awarded by the Space Telescope Science Institute. 

(Some of) The data products presented herein were retrieved from the Dawn JWST Archive (DJA). DJA is an initiative of the Cosmic Dawn Center (DAWN), which is funded by the Danish National Research Foundation under grant DNRF140.

DN was supported by grants ATP-21-0013, ATP-23-0002, and ADSPS-23-0007. DN additionally thanks the Aspen Center for Physics which is supported by National Science Foundation grant PHY-1607611, which is where the original framework for the {\sc powderday} code base was developed.

PT acknowledges support from the National Science Foundation under Cooperative Agreement 2421782 and the Simons Foundation grant MPS-AI-00010515 awarded to NSF-Simons AI Institute for Cosmic Origins -- CosmicAI, \href{https://www.cosmicai.org/}{https://www.cosmicai.org/}.

J.B.M.~was supported by the National Science Foundation under Grants AST-2307354 and AST-2408637, and NASA through grant JWST-GO-03224. 

HL is supported by the National Key R\&D Program of China No. 2023YFB3002502, the National Natural Science Foundation of China under No. 12373006 and 12533004, and the China Manned Space Program with grant No. CMS-CSST-2025-A10.

}
\end{acknowledgments}

\bibliography{references_merged}

@ARTICLE{Aniano2012,
       author = {{Aniano}, G. and {Draine}, B.~T. and {Calzetti}, D. and {Dale}, D.~A. and {Engelbracht}, C.~W. and {Gordon}, K.~D. and {Hunt}, L.~K. and {Kennicutt}, R.~C. and {Krause}, O. and {Leroy}, A.~K. and {Rix}, H.-W. and {Roussel}, H. and {Sandstrom}, K. and {Sauvage}, M. and {Walter}, F. and {Armus}, L. and {Bolatto}, A.~D. and {Crocker}, A. and {Donovan Meyer}, J. and {Galametz}, M. and {Helou}, G. and {Hinz}, J. and {Johnson}, B.~D. and {Koda}, J. and {Montiel}, E. and {Murphy}, E.~J. and {Skibba}, R. and {Smith}, J.-D.~T. and {Wolfire}, M.~G.},
        title = "{Modeling Dust and Starlight in Galaxies Observed by Spitzer and Herschel: NGC 628 and NGC 6946}",
      journal = {\apj},
         year = 2012,
        month = sep,
       volume = {756},
       number = {2},
          eid = {138},
        pages = {138},
          doi = {10.1088/0004-637X/756/2/138},
archivePrefix = {arXiv},
       eprint = {1207.4186},
 primaryClass = {astro-ph.CO},
       adsurl = {https://ui.adsabs.harvard.edu/abs/2012ApJ...756..138A}
}

@ARTICLE{Bohlin1978,
       author = {{Bohlin}, R.~C. and {Savage}, B.~D. and {Drake}, J.~F.},
        title = "{A survey of interstellar H I from Lalpha absorption measurements. II.}",
      journal = {\apj},
         year = 1978,
        month = aug,
       volume = {224},
        pages = {132-142},
          doi = {10.1086/156357},
       adsurl = {https://ui.adsabs.harvard.edu/abs/1978ApJ...224..132B}
}

@ARTICLE{Allen2025,
       author = {{Allen}, Natalie and {Oesch}, Pascal A. and {Toft}, Sune and {Matharu}, Jasleen and {McPartland}, Conor J.~R. and {Weibel}, Andrea and {Brammer}, Gabe and {Bowler}, Rebecca A.~A. and {Ito}, Kei and {Gottumukkala}, Rashmi and {Rizzo}, Francesca and {Valentino}, Francesco and {Varadaraj}, Rohan G. and {Weaver}, John R. and {Whitaker}, Katherine E.},
        title = "{Galaxy size and mass build-up in the first 2 Gyr of cosmic history from multi-wavelength JWST NIRCam imaging}",
      journal = {\aap},
         year = 2025,
        month = jun,
       volume = {698},
          eid = {A30},
        pages = {A30},
          doi = {10.1051/0004-6361/202452690},
archivePrefix = {arXiv},
       eprint = {2410.16354},
 primaryClass = {astro-ph.GA},
       adsurl = {https://ui.adsabs.harvard.edu/abs/2025A&A...698A..30A}
}

@ARTICLE{Austin2024,
       author = {{Austin}, Duncan and {Conselice}, Christopher J. and {Adams}, Nathan J. and {Harvey}, Thomas and {Duan}, Qiao and {Trussler}, James and {Li}, Qiong and {Juodzbalis}, Ignas and {Ormerod}, Katherine and {Ferreira}, Leonardo and {Westcott}, Lewi and {Harris}, Honor and {Wilkins}, Stephen M. and {Bhatawdekar}, Rachana and {Caruana}, Joseph and {Coe}, Dan and {Cohen}, Seth H. and {Driver}, Simon P. and {D'Silva}, Jordan C.~J. and {Frye}, Brenda and {Furtak}, Lukas J. and {Grogin}, Norman A. and {Hathi}, Nimish P. and {Holwerda}, Benne W. and {Jansen}, Rolf A. and {Koekemoer}, Anton M. and {Marshall}, Madeline A. and {Nonino}, Mario and {Ortiz}, III, Rafael and {Pirzkal}, Nor and {Robotham}, Aaron and {Ryan}, Jr., Russell E. and {Summers}, Jake and {Willmer}, Christopher N.~A. and {Windhorst}, Rogier A. and {Yan}, Haojing and {Zackrisson}, Erik},
        title = "{EPOCHS III: Unbiased UV continuum slopes at 6.5<z<13 from combined PEARLS GTO and public JWST NIRCam imaging}",
      journal = {arXiv e-prints},
         year = 2024,
        month = apr,
          eid = {arXiv:2404.10751},
        pages = {arXiv:2404.10751},
          doi = {10.48550/arXiv.2404.10751},
archivePrefix = {arXiv},
       eprint = {2404.10751},
 primaryClass = {astro-ph.GA},
       adsurl = {https://ui.adsabs.harvard.edu/abs/2024arXiv240410751A}
}

@article{bakx_accurate_2021,
    title = {Accurate dust temperature determination in a z = 7.13 galaxy},
    volume = {508},
    doi = {10.1093/mnrasl/slab104},
    number = {1},
    journal = {\mnras},
    author = {Bakx, Tom J. L. C. and Sommovigo, Laura and Carniani, Stefano and Ferrara, Andrea and Akins, Hollis B. and Fujimoto, Seiji and Hagimoto, Masato and Knudsen, Kirsten K. and Pallottini, Andrea and Tamura, Yoichi and Watson, Darach},
    month = nov,
    year = {2021},
    pages = {L58--L63},
}

@ARTICLE{Boquien2019,
       author = {{Boquien}, M. and {Burgarella}, D. and {Roehlly}, Y. and {Buat}, V. and {Ciesla}, L. and {Corre}, D. and {Inoue}, A.~K. and {Salas}, H.},
        title = "{CIGALE: a python Code Investigating GALaxy Emission}",
      journal = {\aap},
         year = 2019,
        month = feb,
       volume = {622},
          eid = {A103},
        pages = {A103},
          doi = {10.1051/0004-6361/201834156},
archivePrefix = {arXiv},
       eprint = {1811.03094},
 primaryClass = {astro-ph.GA},
       adsurl = {https://ui.adsabs.harvard.edu/abs/2019A&A...622A.103B}
}

@article{calzetti_dust_2000,
    title = {The {Dust} {Content} and {Opacity} of {Actively} {Star}-forming {Galaxies}},
    volume = {533},
    issn = {0004-637X},
    url = {https://ui.adsabs.harvard.edu/abs/2000ApJ...533..682C},
    doi = {10.1086/308692},
    urldate = {2022-08-30},
    journal = {The Astrophysical Journal},
    author = {Calzetti, Daniela and Armus, Lee and Bohlin, Ralph C. and Kinney, Anne L. and Koornneef, Jan and Storchi-Bergmann, Thaisa},
    month = apr,
    year = {2000},
    note = {ADS Bibcode: 2000ApJ...533..682C},
    pages = {682--695},
}

@ARTICLE{Casey2022,
       author = {{Casey}, Caitlin M. and {Kartaltepe}, Jeyhan S. and {Drakos}, Nicole E. and {Franco}, Maximilien and {Ilbert}, Olivier and {Rose}, Caitlin and {Cox}, Isabella G. and {Nightingale}, James W. and {Robertson}, Brant E. and {Silverman}, John D. and {Koekemoer}, Anton M. and {Massey}, Richard and {McCracken}, Henry Joy and {Rhodes}, Jason and {Akins}, Hollis B. and {Amvrosiadis}, Aristeidis and {Arango-Toro}, Rafael C. and {Bagley}, Micaela B. and {Capak}, Peter L. and {Champagne}, Jaclyn B. and {Chartab}, Nima and {Chavez Ortiz}, Oscar A. and {Cooke}, Kevin C. and {Cooper}, Olivia R. and {Darvish}, Behnam and {Ding}, Xuheng and {Faisst}, Andreas L. and {Finkelstein}, Steven L. and {Fujimoto}, Seiji and {Gentile}, Fabrizio and {Gillman}, Steven and {Gould}, Katriona M.~L. and {Gozaliasl}, Ghassem and {Harish}, Santosh and {Hayward}, Christopher C. and {He}, Qiuhan and {Hemmati}, Shoubaneh and {Hirschmann}, Michaela and {Jin}, Shuowen and {Khostovan}, Ali Ahmad and {Kokorev}, Vasily and {Lambrides}, Erini and {Laigle}, Clotilde and {Leung}, Gene C.~K. and {Liu}, Daizhong and {Liaudat}, Tobias and {Long}, Arianna S. and {Magdis}, Georgios and {Mahler}, Guillaume and {Mainieri}, Vincenzo and {Manning}, Sinclaire M. and {Maraston}, Claudia and {Martin}, Crystal L. and {McCleary}, Jacqueline E. and {McKinney}, Jed and {McPartland}, Conor J.~R. and {Mobasher}, Bahram and {Pattnaik}, Rohan and {Renzini}, Alvio and {Rich}, R. Michael and {Sanders}, David B. and {Sattari}, Zahra and {Scognamiglio}, Diana and {Scoville}, Nick and {Sheth}, Kartik and {Shuntov}, Marko and {Sparre}, Martin and {Suzuki}, Tomoko L. and {Talia}, Margherita and {Toft}, Sune and {Trakhtenbrot}, Benny and {Urry}, C. Megan and {Valentino}, Francesco and {Vanderhoof}, Brittany N. and {Vardoulaki}, Eleni and {Weaver}, John R. and {Whitaker}, Katherine E. and {Wilkins}, Stephen M. and {Yang}, Lilan and {Zavala}, Jorge A.},
        title = "{COSMOS-Web: An Overview of the JWST Cosmic Origins Survey}",
      journal = {arXiv e-prints},
         year = 2023,
        month = nov,
          eid = {arXiv:2211.07865},
        pages = {arXiv:2211.07865},
archivePrefix = {arXiv},
       eprint = {2211.07865},
 primaryClass = {astro-ph.GA},
       adsurl = {https://ui.adsabs.harvard.edu/abs/2022arXiv221107865C}
}

@ARTICLE{Casey2024,
       author = {{Casey}, Caitlin M. and {Akins}, Hollis B. and {Shuntov}, Marko and {Ilbert}, Olivier and {Paquereau}, Louise and {Franco}, Maximilien and {Hayward}, Christopher C. and {Finkelstein}, Steven L. and {Boylan-Kolchin}, Michael and {Robertson}, Brant E. and {Allen}, Natalie and {Brinch}, Malte and {Cooper}, Olivia R. and {Ding}, Xuheng and {Drakos}, Nicole E. and {Faisst}, Andreas L. and {Fujimoto}, Seiji and {Gillman}, Steven and {Harish}, Santosh and {Hirschmann}, Michaela and {Jin}, Shuowen and {Kartaltepe}, Jeyhan S. and {Koekemoer}, Anton M. and {Kokorev}, Vasily and {Liu}, Daizhong and {Long}, Arianna S. and {Magdis}, Georgios and {Maraston}, Claudia and {Martin}, Crystal L. and {McCracken}, Henry Joy and {McKinney}, Jed and {Mobasher}, Bahram and {Rhodes}, Jason and {Rich}, R. Michael and {Sanders}, David B. and {Silverman}, John D. and {Toft}, Sune and {Vijayan}, Aswin P. and {Weaver}, John R. and {Wilkins}, Stephen M. and {Yang}, Lilan and {Zavala}, Jorge A.},
        title = "{COSMOS-Web: Intrinsically Luminous z {\ensuremath{\gtrsim}} 10 Galaxy Candidates Test Early Stellar Mass Assembly}",
      journal = {\apj},
         year = 2024,
        month = apr,
       volume = {965},
       number = {1},
          eid = {98},
        pages = {98},
          doi = {10.3847/1538-4357/ad2075},
archivePrefix = {arXiv},
       eprint = {2308.10932},
 primaryClass = {astro-ph.GA},
       adsurl = {https://ui.adsabs.harvard.edu/abs/2024ApJ...965...98C}
}

@ARTICLE{Casey2024lrds,
       author = {{Casey}, Caitlin M. and {Akins}, Hollis B. and {Kokorev}, Vasily and {McKinney}, Jed and {Cooper}, Olivia R. and {Long}, Arianna S. and {Franco}, Maximilien and {Manning}, Sinclaire M.},
        title = "{Dust in Little Red Dots}",
      journal = {\apjl},
         year = 2024,
        month = nov,
       volume = {975},
       number = {1},
          eid = {L4},
        pages = {L4},
          doi = {10.3847/2041-8213/ad7ba7},
archivePrefix = {arXiv},
       eprint = {2407.05094},
 primaryClass = {astro-ph.GA},
       adsurl = {https://ui.adsabs.harvard.edu/abs/2024ApJ...975L...4C}
}

@article{casey_dust_2026,
    title = {Dust in the {Average} {Galaxy}: {Attenuation}, {Emission}, and {Opacity} from 0{\textless}z{\textless}7},
    doi = {10.48550/arXiv.2606.17270},
    journal = {arXiv e-prints},
    author = {Casey, Caitlin M. and Akins, Hollis B. and Battisti, Andrew J. and McKinney, Jed and Treister, Ezequiel and Zavala, Jorge A. and Algera, Hiddo and Aravena, Manuel and Cheng, Yingjie and Drakos, Nicole E. and Faisst, Andreas L. and Franco, Maximilien and Fujimoto, Seiji and Gozaliasl, Ghassem and Hadi, Ali and Harish, Santosh and Hirschmann, Michaela and Ilbert, Olivier and Inayoshi, Kohei and Kartaltepe, Jeyhan S. and Koekemoer, Anton M. and Lagos, Claudia del P. and Lambrides, Erini and Laishram, Ronaldo and Liu, Daizhong and Long, Arianna S. and Magdis, Georgios E. and Manning, Sinclaire M. and Martin, Crystal L. and Martinez, III, Felix and Massey, Richard and McCleary, Jacqueline E. and McCracken, Henry Joy and Moscardini, Lauro and Narayanan, Desika and Paquereau, Louise and Rhodes, Jason and Robertson, Brant E. and Samir, Rasha M. and Scarlata, Claudia and Shuntov, Marko and Sommovigo, Laura and Vijayan, Aswin P. and Wang, Wuji and Xu, Can and Zimmerman, Dhruv},
    month = jun,
    year = {2026},
    pages = {arXiv:2606.17270},
}

@article{conroy_modeling_2013,
    title = {Modeling the {Panchromatic} {Spectral} {Energy} {Distributions} of {Galaxies}},
    volume = {51},
    issn = {0066-4146},
    url = {https://ui.adsabs.harvard.edu/abs/2013ARA&A..51..393C/abstract},
    doi = {10.1146/annurev-astro-082812-141017},
    language = {en},
    number = {1},
    urldate = {2025-12-11},
    journal = {Annual Review of Astronomy and Astrophysics},
    author = {Conroy, Charlie},
    month = aug,
    year = {2013},
    pages = {393--455},
}

@ARTICLE{Conselice2014,
       author = {{Conselice}, Christopher J.},
        title = "{The Evolution of Galaxy Structure Over Cosmic Time}",
      journal = {\araa},
         year = 2014,
        month = aug,
       volume = {52},
        pages = {291-337},
          doi = {10.1146/annurev-astro-081913-040037},
archivePrefix = {arXiv},
       eprint = {1403.2783},
 primaryClass = {astro-ph.GA},
       adsurl = {https://ui.adsabs.harvard.edu/abs/2014ARA&A..52..291C}
}

@ARTICLE{Cullen2024,
       author = {{Cullen}, F. and {McLeod}, D.~J. and {McLure}, R.~J. and {Dunlop}, J.~S. and {Donnan}, C.~T. and {Carnall}, A.~C. and {Keating}, L.~C. and {Magee}, D. and {Arellano-Cordova}, K.~Z. and {Bowler}, R.~A.~A. and {Begley}, R. and {Flury}, S.~R. and {Hamadouche}, M.~L. and {Stanton}, T.~M.},
        title = "{The ultraviolet continuum slopes of high-redshift galaxies: evidence for the emergence of dust-free stellar populations at z > 10}",
      journal = {\mnras},
         year = 2024,
        month = jun,
       volume = {531},
       number = {1},
        pages = {997-1020},
          doi = {10.1093/mnras/stae1211},
archivePrefix = {arXiv},
       eprint = {2311.06209},
 primaryClass = {astro-ph.GA},
       adsurl = {https://ui.adsabs.harvard.edu/abs/2024MNRAS.531..997C}
}

@ARTICLE{Draine2007,
       author = {{Draine}, B.~T. and {Dale}, D.~A. and {Bendo}, G. and {Gordon}, K.~D. and {Smith}, J.~D.~T. and {Armus}, L. and {Engelbracht}, C.~W. and {Helou}, G. and {Kennicutt}, R.~C., Jr. and {Li}, A. and {Roussel}, H. and {Walter}, F. and {Calzetti}, D. and {Moustakas}, J. and {Murphy}, E.~J. and {Rieke}, G.~H. and {Bot}, C. and {Hollenbach}, D.~J. and {Sheth}, K. and {Teplitz}, H.~I.},
        title = "{Dust Masses, PAH Abundances, and Starlight Intensities in the SINGS Galaxy Sample}",
      journal = {\apj},
         year = 2007,
        month = jul,
       volume = {663},
       number = {2},
        pages = {866-894},
          doi = {10.1086/518306},
archivePrefix = {arXiv},
       eprint = {astro-ph/0703213},
 primaryClass = {astro-ph},
       adsurl = {https://ui.adsabs.harvard.edu/abs/2007ApJ...663..866D}
}

@ARTICLE{Draine2014,
       author = {{Draine}, B.~T. and {Aniano}, G. and {Krause}, Oliver and {Groves}, Brent and {Sandstrom}, Karin and {Braun}, Robert and {Leroy}, Adam and {Klaas}, Ulrich and {Linz}, Hendrik and {Rix}, Hans-Walter and {Schinnerer}, Eva and {Schmiedeke}, Anika and {Walter}, Fabian},
        title = "{Andromeda's Dust}",
      journal = {\apj},
         year = 2014,
        month = jan,
       volume = {780},
       number = {2},
          eid = {172},
        pages = {172},
          doi = {10.1088/0004-637X/780/2/172},
archivePrefix = {arXiv},
       eprint = {1306.2304},
 primaryClass = {astro-ph.CO},
       adsurl = {https://ui.adsabs.harvard.edu/abs/2014ApJ...780..172D}
}

@article{endsley_radio_2022,
    title = {Radio and far-{IR} emission associated with a massive star-forming galaxy candidate at z ≃ 6.8: a radio-loud {AGN} in the reionization era?},
    volume = {512},
    doi = {10.1093/mnras/stac737},
    number = {3},
    journal = {\mnras},
    author = {Endsley, Ryan and Stark, Daniel P. and Fan, Xiaohui and Smit, Renske and Wang, Feige and Yang, Jinyi and Hainline, Kevin and Lyu, Jianwei and Bouwens, Rychard and Schouws, Sander},
    month = may,
    year = {2022},
    pages = {4248--4261},
}

@article{Ferrara2022,
    title = {The {ALMA} {REBELS} {Survey}. {Epoch} of {Reionization} giants: {Properties} of dusty galaxies at z ≍ 7},
    volume = {512},
    issn = {0035-8711},
    shorttitle = {The {ALMA} {REBELS} {Survey}. {Epoch} of {Reionization} giants},
    url = {https://ui.adsabs.harvard.edu/abs/2022MNRAS.512...58F},
    doi = {10.1093/mnras/stac460},
    urldate = {2022-09-01},
    journal = {Monthly Notices of the Royal Astronomical Society},
    author = {Ferrara, A. and Sommovigo, L. and Dayal, P. and Pallottini, A. and Bouwens, R. J. and Gonzalez, V. and Inami, H. and Smit, R. and Bowler, R. A. A. and Endsley, R. and Oesch, P. and Schouws, S. and Stark, D. and Stefanon, M. and Aravena, M. and da Cunha, E. and De Looze, I. and Fudamoto, Y. and Graziani, L. and Hodge, J. and Riechers, D. and Schneider, R. and Algera, H. S. B. and Barrufet, L. and Hygate, A. P. S. and Labbé, I. and Li, C. and Nanayakkara, T. and Topping, M. and van der Werf, P.},
    month = may,
    year = {2022},
    note = {ADS Bibcode: 2022MNRAS.512...58F},
    pages = {58--72},
}

@ARTICLE{Ferrara2024bm,
       author = {{Ferrara}, A. and {Pallottini}, A. and {Sommovigo}, L.},
        title = "{Blue Monsters at $z>10$. Where has all their dust gone?}",
      journal = {arXiv e-prints},
         year = 2024,
        month = oct,
          eid = {arXiv:2410.19042},
        pages = {arXiv:2410.19042},
          doi = {10.48550/arXiv.2410.19042},
archivePrefix = {arXiv},
       eprint = {2410.19042},
 primaryClass = {astro-ph.GA},
       adsurl = {https://ui.adsabs.harvard.edu/abs/2024arXiv241019042F}
}

@article{Fudamoto2021,
    title = {Normal, dust-obscured galaxies in the epoch of reionization},
    volume = {597},
    issn = {0028-0836},
    url = {https://ui.adsabs.harvard.edu/abs/2021Natur.597..489F},
    doi = {10.1038/s41586-021-03846-z},
    urldate = {2022-07-05},
    journal = {Nature},
    author = {Fudamoto, Y. and Oesch, P. A. and Schouws, S. and Stefanon, M. and Smit, R. and Bouwens, R. J. and Bowler, R. A. A. and Endsley, R. and Gonzalez, V. and Inami, H. and Labbe, I. and Stark, D. and Aravena, M. and Barrufet, L. and da Cunha, E. and Dayal, P. and Ferrara, A. and Graziani, L. and Hodge, J. and Hutter, A. and Li, Y. and De Looze, I. and Nanayakkara, T. and Pallottini, A. and Riechers, D. and Schneider, R. and Ucci, G. and van der Werf, P. and White, C.},
    month = sep,
    year = {2021},
    note = {ADS Bibcode: 2021Natur.597..489F},
    pages = {489--492},
}

@ARTICLE{Hashimoto2018,
       author = {{Hashimoto}, Takuya and {Inoue}, Akio K. and {Tamura}, Yoichi and
         {Matsuo}, Hiroshi and {Mawatari}, Ken and {Yamaguchi}, Yuki},
        title = "{Detections of [OIII] 88 $\mu$m in Two Quasars in the Reionization Epoch}",
      journal = {arXiv e-prints},
         year = "2018",
        month = "Oct",
          eid = {arXiv:1811.00030},
        pages = {arXiv:1811.00030},
archivePrefix = {arXiv},
       eprint = {1811.00030},
 primaryClass = {astro-ph.GA},
       adsurl = {https://ui.adsabs.harvard.edu/abs/2018arXiv181100030H}
}

@ARTICLE{Hashimoto2019,
       author = {{Hashimoto}, Takuya and {Inoue}, Akio K. and {Mawatari}, Ken and
         {Tamura}, Yoichi and {Matsuo}, Hiroshi and {Furusawa}, Hisanori and
         {Harikane}, Yuichi and {Shibuya}, Takatoshi and {Knudsen}, Kirsten K. and
         {Kohno}, Kotaro and {Ono}, Yoshiaki and {Zackrisson}, Erik and
         {Okamoto}, Takashi and {Kashikawa}, Nobunari and {Oesch}, Pascal A. and
         {Ouchi}, Masami and {Ota}, Kazuaki and {Shimizu}, Ikkoh and
         {Taniguchi}, Yoshiaki and {Umehata}, Hideki and {Watson}, Darach},
        title = "{Big Three Dragons: A z = 7.15 Lyman-break galaxy detected in [O III] 88 {\ensuremath{\mu}}m, [C II] 158 {\ensuremath{\mu}}m, and dust continuum with ALMA}",
      journal = {\pasj},
         year = "2019",
        month = "Aug",
       volume = {71},
       number = {4},
          eid = {71},
        pages = {71},
          doi = {10.1093/pasj/psz049},
archivePrefix = {arXiv},
       eprint = {1806.00486},
 primaryClass = {astro-ph.GA},
       adsurl = {https://ui.adsabs.harvard.edu/abs/2019PASJ...71...71H}
}

@ARTICLE{Hughes1998,
       author = {{Hughes}, David H. and {Serjeant}, Stephen and {Dunlop}, James and {Rowan-Robinson}, Michael and {Blain}, Andrew and {Mann}, Robert G. and {Ivison}, Rob and {Peacock}, John and {Efstathiou}, Andreas and {Gear}, Walter and {Oliver}, Seb and {Lawrence}, Andy and {Longair}, Malcolm and {Goldschmidt}, Pippa and {Jenness}, Tim},
        title = "{High-redshift star formation in the Hubble Deep Field revealed by a submillimetre-wavelength survey}",
      journal = {\nat},
         year = 1998,
        month = jul,
       volume = {394},
       number = {6690},
        pages = {241-247},
          doi = {10.1038/28328},
archivePrefix = {arXiv},
       eprint = {astro-ph/9806297},
 primaryClass = {astro-ph},
       adsurl = {https://ui.adsabs.harvard.edu/abs/1998Natur.394..241H}
}

@article{marinacci_simulating_2019,
    title = {Simulating the interstellar medium and stellar feedback on a moving mesh: implementation and isolated galaxies},
    volume = {489},
    issn = {0035-8711},
    shorttitle = {Simulating the interstellar medium and stellar feedback on a moving mesh},
    url = {https://ui.adsabs.harvard.edu/abs/2019MNRAS.489.4233M/abstract},
    doi = {10.1093/mnras/stz2391},
    language = {en},
    number = {3},
    urldate = {2026-03-19},
    journal = {Monthly Notices of the Royal Astronomical Society},
    author = {Marinacci, Federico and Sales, Laura V. and Vogelsberger, Mark and Torrey, Paul and Springel, Volker},
    month = nov,
    year = {2019},
    pages = {4233--4260},
}

@ARTICLE{Markov2024,
       author = {{Markov}, Vladan and {Gallerani}, Simona and {Ferrara}, Andrea and {Pallottini}, Andrea and {Parlanti}, Eleonora and {Mascia}, Fabio Di and {Sommovigo}, Laura and {Kohandel}, Mahsa},
        title = "{The evolution of dust attenuation in z {\ensuremath{\approx}} 2-12 galaxies observed by JWST}",
      journal = {Nature Astronomy},
         year = 2024,
        month = nov,
          doi = {10.1038/s41550-024-02426-1},
archivePrefix = {arXiv},
       eprint = {2402.05996},
 primaryClass = {astro-ph.GA},
       adsurl = {https://ui.adsabs.harvard.edu/abs/2024NatAs.tmp...20M}
}

@ARTICLE{Diemer2018,
       author = {{Diemer}, Benedikt},
        title = "{COLOSSUS: A Python Toolkit for Cosmology, Large-scale Structure, and Dark Matter Halos}",
      journal = {\apjs},
         year = 2018,
        month = dec,
       volume = {239},
       number = {2},
          eid = {35},
        pages = {35},
          doi = {10.3847/1538-4365/aaee8c},
archivePrefix = {arXiv},
       eprint = {1712.04512},
 primaryClass = {astro-ph.CO},
       adsurl = {https://ui.adsabs.harvard.edu/abs/2018ApJS..239...35D}
}

@ARTICLE{Watson2013,
       author = {{Watson}, William A. and {Iliev}, Ilian T. and {D'Aloisio}, Anson and {Knebe}, Alexander and {Shapiro}, Paul R. and {Yepes}, Gustavo},
        title = "{The halo mass function through the cosmic ages}",
      journal = {\mnras},
         year = 2013,
        month = aug,
       volume = {433},
       number = {2},
        pages = {1230-1245},
          doi = {10.1093/mnras/stt791},
archivePrefix = {arXiv},
       eprint = {1212.0095},
 primaryClass = {astro-ph.CO},
       adsurl = {https://ui.adsabs.harvard.edu/abs/2013MNRAS.433.1230W}
}

@ARTICLE{Mosenkov2019,
       author = {{Mosenkov}, A.~V. and {Baes}, M. and {Bianchi}, S. and {Casasola}, V. and {Cassar{\`a}}, L.~P. and {Clark}, C.~J.~R. and {Davies}, J. and {De Looze}, I. and {De Vis}, P. and {Fritz}, J. and {Galametz}, M. and {Galliano}, F. and {Jones}, A.~P. and {Lianou}, S. and {Madden}, S.~C. and {Nersesian}, A. and {Smith}, M.~W.~L. and {Tr{\v{c}}ka}, A. and {Verstocken}, S. and {Viaene}, S. and {Vika}, M. and {Xilouris}, E.},
        title = "{Dust emission profiles of DustPedia galaxies}",
      journal = {\aap},
         year = 2019,
        month = feb,
       volume = {622},
          eid = {A132},
        pages = {A132},
          doi = {10.1051/0004-6361/201833932},
archivePrefix = {arXiv},
       eprint = {1811.08923},
 primaryClass = {astro-ph.GA},
       adsurl = {https://ui.adsabs.harvard.edu/abs/2019A&A...622A.132M}
}

@article{mckinney_modeling_2025,
    title = {Modeling {Galaxies} in the {Early} {Universe} with {Supernova} {Dust} {Attenuation}},
    volume = {985},
    issn = {0004-637X},
    url = {https://ui.adsabs.harvard.edu/abs/2025ApJ...985L..21M/abstract},
    doi = {10.3847/2041-8213/add15d},
    language = {en},
    number = {1},
    urldate = {2025-12-11},
    journal = {The Astrophysical Journal},
    author = {McKinney, Jed and Cooper, Olivia R. and Casey, Caitlin M. and Muñoz, Julian B. and Akins, Hollis and Lambrides, Erini and Long, Arianna S.},
    month = may,
    year = {2025},
    pages = {L21},
}

@article{mckinnon_dust_2016,
    title = {Dust formation in {Milky} {Way}-like galaxies},
    volume = {457},
    issn = {0035-8711},
    url = {https://ui.adsabs.harvard.edu/abs/2016MNRAS.457.3775M/abstract},
    doi = {10.1093/mnras/stw253},
    language = {en},
    number = {4},
    urldate = {2026-03-19},
    journal = {Monthly Notices of the Royal Astronomical Society},
    author = {McKinnon, Ryan and Torrey, Paul and Vogelsberger, Mark},
    month = apr,
    year = {2016},
    pages = {3775--3800},
}

@article{mckinnon_simulating_2017,
    title = {Simulating the dust content of galaxies: successes and failures},
    volume = {468},
    issn = {0035-8711},
    shorttitle = {Simulating the dust content of galaxies},
    url = {https://ui.adsabs.harvard.edu/abs/2017MNRAS.468.1505M/abstract},
    doi = {10.1093/mnras/stx467},
    language = {en},
    number = {2},
    urldate = {2026-03-19},
    journal = {Monthly Notices of the Royal Astronomical Society},
    author = {McKinnon, Ryan and Torrey, Paul and Vogelsberger, Mark and Hayward, Christopher C. and Marinacci, Federico},
    month = jun,
    year = {2017},
    pages = {1505--1521},
}

@article{mckinnon_simulating_2018,
    title = {Simulating galactic dust grain evolution on a moving mesh},
    volume = {478},
    issn = {0035-8711},
    url = {https://ui.adsabs.harvard.edu/abs/2018MNRAS.478.2851M/abstract},
    doi = {10.1093/mnras/sty1248},
    language = {en},
    number = {3},
    urldate = {2026-03-19},
    journal = {Monthly Notices of the Royal Astronomical Society},
    author = {McKinnon, Ryan and Vogelsberger, Mark and Torrey, Paul and Marinacci, Federico and Kannan, Rahul},
    month = aug,
    year = {2018},
    pages = {2851--2886},
}

@ARTICLE{Morales2024,
       author = {{Morales}, Alexa M. and {Finkelstein}, Steven L. and {Leung}, Gene C.~K. and {Bagley}, Micaela B. and {Cleri}, Nikko J. and {Dave}, Romeel and {Dickinson}, Mark and {Ferguson}, Henry C. and {Hathi}, Nimish P. and {Jones}, Ewan and {Koekemoer}, Anton M. and {Papovich}, Casey and {P{\'e}rez-Gonz{\'a}lez}, Pablo G. and {Pirzkal}, Nor and {Smith}, Britton and {Wilkins}, Stephen M. and {Yung}, L.~Y. Aaron},
        title = "{Rest-frame UV Colors for Faint Galaxies at z {\ensuremath{\sim}} 9{\textendash}16 with the JWST NGDEEP Survey}",
      journal = {\apjl},
         year = 2024,
        month = apr,
       volume = {964},
       number = {2},
          eid = {L24},
        pages = {L24},
          doi = {10.3847/2041-8213/ad2de4},
archivePrefix = {arXiv},
       eprint = {2311.04294},
 primaryClass = {astro-ph.GA},
       adsurl = {https://ui.adsabs.harvard.edu/abs/2024ApJ...964L..24M}
}

@article{Narayanan2023,
    title = {A {Framework} for {Modeling} {Polycyclic} {Aromatic} {Hydrocarbon} {Emission} in {Galaxy} {Evolution} {Simulations}},
    volume = {951},
    issn = {0004-637X},
    url = {https://ui.adsabs.harvard.edu/abs/2023ApJ...951..100N/abstract},
    doi = {10.3847/1538-4357/accf8d},
    language = {en},
    number = {2},
    urldate = {2026-03-19},
    journal = {The Astrophysical Journal},
    author = {Narayanan, Desika and Smith, J.-D. T. and Hensley, Brandon S. and Li, Qi and Hu, Chia-Yu and Sandstrom, Karin and Torrey, Paul and Vogelsberger, Mark and Marinacci, Federico and Sales, Laura V.},
    month = jul,
    year = {2023},
    pages = {100},
}

@ARTICLE{Narayanan2025,
       author = {{Narayanan}, Desika and {Stark}, Daniel P. and {Finkelstein}, Steven L. and {Torrey}, Paul and {Li}, Qi and {Cullen}, Fergus and {Topping}, Micheal W. and {Marinacci}, Federico and {Sales}, Laura V. and {Shen}, Xuejian and {Vogelsberger}, Mark},
        title = "{The Ultraviolet Slopes of Early Universe Galaxies: The Impact of Bursty Star Formation, Dust, and Nebular Continuum Emission}",
      journal = {\apj},
         year = 2025,
        month = mar,
       volume = {982},
       number = {1},
          eid = {7},
        pages = {7},
          doi = {10.3847/1538-4357/adb41c},
       adsurl = {https://ui.adsabs.harvard.edu/abs/2025ApJ...982....7N}
}

@article{narayanan_powderday_2021,
    title = {{POWDERDAY}: {Dust} {Radiative} {Transfer} for {Galaxy} {Simulations}},
    volume = {252},
    issn = {0067-0049},
    shorttitle = {{POWDERDAY}},
    url = {https://ui.adsabs.harvard.edu/abs/2021ApJS..252...12N},
    doi = {10.3847/1538-4365/abc487},
    urldate = {2022-07-05},
    journal = {The Astrophysical Journal Supplement Series},
    author = {Narayanan, Desika and Turk, Matthew J. and Robitaille, Thomas and Kelly, Ashley J. and McClellan, B. Connor and Sharma, Ray S. and Garg, Prerak and Abruzzo, Matthew and Choi, Ena and Conroy, Charlie and Johnson, Benjamin D. and Kimock, Benjamin and Li, Qi and Lovell, Christopher C. and Lower, Sidney and Privon, George C. and Roberts, Jonathan and Sethuram, Snigdaa and Snyder, Gregory F. and Thompson, Robert and Wise, John H.},
    month = jan,
    year = {2021},
    note = {ADS Bibcode: 2021ApJS..252...12N},
    pages = {12},
}

@ARTICLE{RemyRuyer2014,
       author = {{R{\'e}my-Ruyer}, A. and {Madden}, S.~C. and {Galliano}, F. and {Galametz}, M. and {Takeuchi}, T.~T. and {Asano}, R.~S. and {Zhukovska}, S. and {Lebouteiller}, V. and {Cormier}, D. and {Jones}, A. and {Bocchio}, M. and {Baes}, M. and {Bendo}, G.~J. and {Boquien}, M. and {Boselli}, A. and {DeLooze}, I. and {Doublier-Pritchard}, V. and {Hughes}, T. and {Karczewski}, O. {\L}. and {Spinoglio}, L.},
        title = "{Gas-to-dust mass ratios in local galaxies over a 2 dex metallicity range}",
      journal = {\aap},
         year = 2014,
        month = mar,
       volume = {563},
          eid = {A31},
        pages = {A31},
          doi = {10.1051/0004-6361/201322803},
archivePrefix = {arXiv},
       eprint = {1312.3442},
 primaryClass = {astro-ph.GA},
       adsurl = {https://ui.adsabs.harvard.edu/abs/2014A&A...563A..31R}
}

@article{Salim2020,
    title = {The {Dust} {Attenuation} {Law} in {Galaxies}},
    volume = {58},
    issn = {0066-4146},
    url = {https://ui.adsabs.harvard.edu/abs/2020ARA&A..58..529S/abstract},
    doi = {10.1146/annurev-astro-032620-021933},
    language = {en},
    urldate = {2022-07-05},
    journal = {Annual Review of Astronomy and Astrophysics},
    author = {Salim, Samir and Narayanan, Desika},
    month = aug,
    year = {2020},
    pages = {529--575},
}

@ARTICLE{Schneider2024,
       author = {{Schneider}, Raffaella and {Maiolino}, Roberto},
        title = "{The formation and cosmic evolution of dust in the early Universe: I. Dust sources}",
      journal = {\aapr},
         year = 2024,
        month = apr,
       volume = {32},
       number = {1},
          eid = {2},
        pages = {2},
          doi = {10.1007/s00159-024-00151-2},
archivePrefix = {arXiv},
       eprint = {2310.00053},
 primaryClass = {astro-ph.GA},
       adsurl = {https://ui.adsabs.harvard.edu/abs/2024A&ARv..32....2S}
}

@ARTICLE{Asano2013,
       author = {{Asano}, Ryosuke S. and {Takeuchi}, Tsutomu T. and {Hirashita}, Hiroyuki and {Inoue}, Akio K.},
        title = "{Dust formation history of galaxies: A critical role of metallicity* for the dust mass growth by accreting materials in the interstellar medium}",
      journal = {Earth, Planets and Space},
         year = 2013,
        month = mar,
       volume = {65},
       number = {3},
        pages = {213-222},
          doi = {10.5047/eps.2012.04.014},
archivePrefix = {arXiv},
       eprint = {1206.0817},
 primaryClass = {astro-ph.GA},
       adsurl = {https://ui.adsabs.harvard.edu/abs/2013EP&S...65..213A}
}

@ARTICLE{Choban2025,
       author = {{Choban}, Caleb R. and {Salim}, Samir and {Kere{\v{s}}}, Du{\v{s}}an and {Hayward}, Christopher C. and {Sandstrom}, Karin M.},
        title = "{A dusty dawn: galactic dust buildup at z {\ensuremath{\gtrsim}} 5}",
      journal = {\mnras},
         year = 2025,
        month = feb,
       volume = {537},
       number = {2},
        pages = {1518-1541},
          doi = {10.1093/mnras/staf118},
archivePrefix = {arXiv},
       eprint = {2408.08962},
 primaryClass = {astro-ph.GA},
       adsurl = {https://ui.adsabs.harvard.edu/abs/2025MNRAS.537.1518C}
}

@ARTICLE{Burgarella2026,
       author = {{Burgarella}, Denis and {Bakx}, Tom J.~L.~C. and {Schneider}, Raffaella and {Algera}, Hiddo S.~B. and {Aurin}, Chloe and {Dewachter}, Tim and {Esmerian}, Clarke and {Graziani}, Luca and {Knudsen}, Kirsten and {Otaki}, Koki},
        title = "{From stardust to interstellar grain growth in the first galaxies: a cosmological transition in dust evolution near z \raisebox{-0.5ex}\textasciitilde 8.9}",
      journal = {arXiv e-prints},
         year = 2026,
        month = sep,
          eid = {arXiv:2609.10290},
        pages = {arXiv:2609.10290},
          doi = {10.48550/arXiv.2609.10290},
archivePrefix = {arXiv},
       eprint = {2609.10290},
 primaryClass = {astro-ph.GA},
       adsurl = {https://ui.adsabs.harvard.edu/abs/2026arXiv260910290B}
}

@ARTICLE{Santini2010,
       author = {{Santini}, P. and {Maiolino}, R. and {Magnelli}, B. and {Silva}, L. and {Grazian}, A. and {Altieri}, B. and {Andreani}, P. and {Aussel}, H. and {Berta}, S. and {Bongiovanni}, A. and {Brisbin}, D. and {Calura}, F. and {Cava}, A. and {Cepa}, J. and {Cimatti}, A. and {Daddi}, E. and {Dannerbauer}, H. and {Dominguez-Sanchez}, H. and {Elbaz}, D. and {Fontana}, A. and {F{\"o}rster Schreiber}, N. and {Genzel}, R. and {Granato}, G.~L. and {Gruppioni}, C. and {Lutz}, D. and {Magdis}, G. and {Magliocchetti}, M. and {Matteucci}, F. and {Nordon}, R. and {P{\'e}rez Garcia}, I. and {Poglitsch}, A. and {Popesso}, P. and {Pozzi}, F. and {Riguccini}, L. and {Rodighiero}, G. and {Saintonge}, A. and {Sanchez-Portal}, M. and {Shao}, L. and {Sturm}, E. and {Tacconi}, L. and {Valtchanov}, I.},
        title = "{The dust content of high-z submillimeter galaxies revealed by Herschel}",
      journal = {\aap},
         year = 2010,
        month = jul,
       volume = {518},
          eid = {L154},
        pages = {L154},
          doi = {10.1051/0004-6361/201014748},
archivePrefix = {arXiv},
       eprint = {1005.5678},
 primaryClass = {astro-ph.CO},
       adsurl = {https://ui.adsabs.harvard.edu/abs/2010A&A...518L.154S}
}

@article{schneider_dust_2014,
    title = {Dust production rate of asymptotic giant branch stars in the {Magellanic} {Clouds}},
    volume = {442},
    issn = {0035-8711},
    url = {https://ui.adsabs.harvard.edu/abs/2014MNRAS.442.1440S/abstract},
    doi = {10.1093/mnras/stu861},
    language = {en},
    number = {2},
    urldate = {2026-03-19},
    journal = {Monthly Notices of the Royal Astronomical Society},
    author = {Schneider, Raffaella and Valiante, Rosa and Ventura, Paolo and dell'Agli, Flavia and Di Criscienzo, Marcella and Hirashita, Hiroyuki and Kemper, Francisca},
    month = aug,
    year = {2014},
    pages = {1440--1450},
}

@ARTICLE{Scoville2014,
       author = {{Scoville}, N. and {Aussel}, H. and {Sheth}, K. and {Scott}, K.~S. and {Sanders}, D. and {Ivison}, R. and {Pope}, A. and {Capak}, P. and {Vanden Bout}, P. and {Manohar}, S. and {Kartaltepe}, J. and {Robertson}, B. and {Lilly}, S.},
        title = "{The Evolution of Interstellar Medium Mass Probed by Dust Emission: ALMA Observations at z = 0.3-2}",
      journal = {\apj},
         year = 2014,
        month = mar,
       volume = {783},
       number = {2},
          eid = {84},
        pages = {84},
          doi = {10.1088/0004-637X/783/2/84},
archivePrefix = {arXiv},
       eprint = {1401.2987},
 primaryClass = {astro-ph.GA},
       adsurl = {https://ui.adsabs.harvard.edu/abs/2014ApJ...783...84S}
}

@article{shivaei_diversity_2025,
    title = {Diversity and {Evolution} of {Dust} {Attenuation} {Curves} from {Redshift} z {\textasciitilde} 1 to 9},
    url = {https://arxiv.org/abs/2509.01795v3},
    language = {en},
    urldate = {2026-06-03},
    journal = {arXiv.org},
    author = {Shivaei, Irene and Naidu, Rohan P. and Montero, Francisco Rodriguez and Matsumoto, Kosei and Leja, Joel and Matthee, Jorryt and Johnson, Benjamin D. and Oesch, Pascal A. and Chevallard, Jacopo and Adamo, Angela and Bodansky, Sarah and Bunker, Andrew J. and Paz, Alba Covelo and Di Cesare, Claudia and Egami, Eiichi and Furtak, Lukas J. and Heintz, Kasper E. and Kramarenko, Ivan and Meyer, Romain A. and Reddy, Naveen A. and Rinaldi, Pierluigi and Tacchella, Sandro and Torralba, Alberto and Witstok, Joris and Wozniak, Michael A. and Xiao, Mengyuan},
    month = sep,
    year = {2025},
}

@ARTICLE{Soliman2024,
       author = {{Soliman}, Nadine H. and {Hopkins}, Philip F. and {Grudi{\'c}}, Michael Y.},
        title = "{Thermodynamics of Giant Molecular Clouds: The Effects of Dust Grain Size}",
      journal = {\apj},
         year = 2024,
        month = nov,
       volume = {975},
       number = {2},
          eid = {284},
        pages = {284},
          doi = {10.3847/1538-4357/ad8087},
archivePrefix = {arXiv},
       eprint = {2407.09343},
 primaryClass = {astro-ph.GA},
       adsurl = {https://ui.adsabs.harvard.edu/abs/2024ApJ...975..284S}
}

@ARTICLE{Katz2025,
       author = {{Katz}, Harley and {Cameron}, Alex J. and {Saxena}, Aayush and {Barrufet}, Laia and {Choustikov}, Nichloas and {Cleri}, Nikko J. and {de Graff}, Anna and {Ellis}, Richard S. and {Fosbury}, Robert A.~E. and {Heintz}, Kasper E. and {Maseda}, Michael and {Matthee}, Jorryt and {McConachie}, Ian and {Oesch}, Pascal A.},
        title = "{21 Balmer Jump Street: The Nebular Continuum at High Redshift and Implications for the Bright Galaxy Problem, UV Continuum Slopes, and Early Stellar Populations}",
      journal = {The Open Journal of Astrophysics},
         year = 2025,
        month = jul,
       volume = {8},
          eid = {104},
        pages = {104},
          doi = {10.33232/001c.142570},
archivePrefix = {arXiv},
       eprint = {2408.03189},
 primaryClass = {astro-ph.GA},
       adsurl = {https://ui.adsabs.harvard.edu/abs/2025OJAp....8E.104K}
}

@ARTICLE{Tamura2019,
       author = {{Tamura}, Yoichi and {Mawatari}, Ken and {Hashimoto}, Takuya and {Inoue}, Akio K. and {Zackrisson}, Erik and {Christensen}, Lise and {Binggeli}, Christian and {Matsuda}, Yuichi and {Matsuo}, Hiroshi and {Takeuchi}, Tsutomu T. and {Asano}, Ryosuke S. and {Sunaga}, Kaho and {Shimizu}, Ikkoh and {Okamoto}, Takashi and {Yoshida}, Naoki and {Lee}, Minju M. and {Shibuya}, Takatoshi and {Taniguchi}, Yoshiaki and {Umehata}, Hideki and {Hatsukade}, Bunyo and {Kohno}, Kotaro and {Ota}, Kazuaki},
        title = "{Detection of the Far-infrared [O III] and Dust Emission in a Galaxy at Redshift 8.312: Early Metal Enrichment in the Heart of the Reionization Era}",
      journal = {\apj},
         year = 2019,
        month = mar,
       volume = {874},
       number = {1},
          eid = {27},
        pages = {27},
          doi = {10.3847/1538-4357/ab0374},
archivePrefix = {arXiv},
       eprint = {1806.04132},
 primaryClass = {astro-ph.GA},
       adsurl = {https://ui.adsabs.harvard.edu/abs/2019ApJ...874...27T}
}

@ARTICLE{Tielens1985,
       author = {{Tielens}, A.~G.~G.~M. and {Hollenbach}, D.},
        title = "{Photodissociation regions. II. A model for the Orion photodissociation region.}",
      journal = {\apj},
         year = 1985,
        month = apr,
       volume = {291},
        pages = {747-754},
          doi = {10.1086/163112},
       adsurl = {https://ui.adsabs.harvard.edu/abs/1985ApJ...291..747T}
}

@ARTICLE{Topping2022,
       author = {{Topping}, Michael W. and {Stark}, Daniel P. and {Endsley}, Ryan and {Plat}, Adele and {Whitler}, Lily and {Chen}, Zuyi and {Charlot}, St{\'e}phane},
        title = "{Searching for Extremely Blue UV Continuum Slopes at $z=7-11$ in JWST/NIRCam Imaging: Implications for Stellar Metallicity and Ionizing Photon Escape in Early Galaxies}",
      journal = {arXiv e-prints},
         year = 2022,
        month = aug,
          eid = {arXiv:2208.01610},
        pages = {arXiv:2208.01610},
archivePrefix = {arXiv},
       eprint = {2208.01610},
 primaryClass = {astro-ph.GA},
       adsurl = {https://ui.adsabs.harvard.edu/abs/2022arXiv220801610T}
}
\bibliographystyle{aasjournal}

\end{document}